\documentclass[preprint,12pt]{elsarticle}

\usepackage{amssymb}

\journal{Nuclear Physics B}

\begin{document}

\begin{frontmatter}



\title{Generative Artificial Intelligence for Software Engineering - A Research Agenda}


\author[inst1,inst2]{Anh Nguyen-Duc}
\author[inst3]{Beatriz Cabrero-Daniel}
\author[inst4]{Adam Przybylek}
\author[inst5]{Chetan Arora}
\author[inst6]{Dron Khanna}
\author[inst7]{Tomas Herda}
\author[inst6]{Usman Rafiq}
\author[inst6]{Jorge Melegati}
\author[inst6]{Eduardo Guerra}
\author[inst8]{Kai-Kristian Kemell}
\author[inst9]{Mika Saari}
\author[inst9]{Zheying Zhang}
\author[inst10,inst11]{Huy Le}
\author[inst10,inst11]{Tho Quan}
\author[inst9]{Pekka Abrahamsson}

\affiliation[inst1]{
  organization={University of South Eastern Norway},
  city={B\o I Telemark}, 
  country={Norway},
  postcode={3800}
}
\affiliation[inst2]{
  organization={Norwegian University of Science and Technology},
  city={Trondheim}, 
  country={Norway},
  postcode={7012}
}
\affiliation[inst3]{
  organization={University of Gothenburg},
  city={Gothenburg}, 
  country={Sweden}
}
\affiliation[inst4]{
  organization={Gdansk University of Technology},
  city={Gdansk}, 
  country={Poland}
}
\affiliation[inst5]{
  organization={Monash University},
  city={Melbourne}, 
  country={Australia}
}
\affiliation[inst6]{
  organization={Free University of Bozen-Bolzano},
  city={Bolzano}, 
  country={Italy}
}
\affiliation[inst7]{
  organization={Austrian Post - Konzern IT},
  city={Vienna}, 
  country={Austria}
}

\affiliation[inst8]{
  organization={University of Helsinki},
  city={Helsinki}, 
  country={Finland}
}

\affiliation[inst9]{
  organization={Tampere University},
  city={Tampere}, 
  country={Finland}
}

\affiliation[inst10]{
  organization={Vietnam National University Ho Chi Minh City},
  city={Hochiminh City}, 
  country={Vietnam}
}

\affiliation[inst11]{
  organization={Ho Chi Minh City University of Technology},
  city={Hochiminh City}, 
  country={Vietnam}
}

\begin{abstract}
  \textbf{Context} Generative Artificial Intelligence (GenAI) tools have become increasingly prevalent in software development, offering assistance to various managerial and technical project activities. Notable examples of these tools include OpenAI’s ChatGPT, GitHub Copilot, and Amazon CodeWhisperer.
  
  \textbf{Objective.} Although many recent publications have explored and evaluated the application of GenAI, a comprehensive understanding of the current development, applications, limitations, and open challenges remains unclear to many. Particularly, we do not have an overall picture of the current state of GenAI technology in practical software engineering usage scenarios.
  
  \textbf{Method.} We conducted a literature review and focus groups for a duration of five months to develop a research agenda on GenAI for Software Engineering. 
  
  \textbf{Results.} We identified 78 open Research Questions (RQs) in 11 areas of Software Engineering. Our results show that it is possible to explore the adoption of GenAI in partial automation and support decision-making in all software development activities. While the current literature is skewed toward software implementation, quality assurance and software maintenance, other areas, such as requirements engineering, software design, and software engineering education, would need further research attention. Common considerations when implementing GenAI include industry-level assessment, dependability and accuracy, data accessibility, transparency, and sustainability aspects associated with the technology.
  
  \textbf{Conclusions.} GenAI is bringing significant changes to the field of software engineering. Nevertheless, the state of research on the topic still remains immature. We believe that this research agenda holds significance and practical value for informing both researchers and practitioners about current applications and guiding future research.
\end{abstract}




\end{frontmatter}


\section{Introduction}
In recent years, Generative Artificial Intelligence (GenAI) has attracted significant attention and interest from Software Engineering~(SE) research and practice. GenAI tools such as GitHub Copilot\footnote{https://github.com/features/copilot} and ChatGPT\footnote{https://openai.com/chatgpt} have been rapidly explored in various professional contexts given their remarkable performance in producing human-like content. The emerging adoption of these tools reintroduces a host of classic concerns about productivity and quality when adopting new technologies. Clearly, code generation and test case optimization are SE tasks that directly benefit from recent large language models (LLMs) \cite{poldrack_ai-assisted_2023,denny_conversing_2022,doderlein_piloting_2023,dong_self-collaboration_2023}. Beyond traditional research on applied Machine Learning (ML), GenAI tools introduced usability and accessibility, leveraging AI-generated content to a broader range of professionals, requiring less technical competence to work and integrate them into existing work environment. GenAI tools are also showing their potential in non-coding tasks, such as assisting requirements engineering, software processes and project management. 

At the moment, GenAI is an active research areas with several challenges and open questions. To do well on specific tasks, LLMs require fine-tuning or training. Research has much focused so far on achieving reliable and scalable GenAI output for different SE tasks. GenAI models are inherently nondeterministic: the same prompt produces different answers on different inference executions \cite{ouyang_llm_2023}. Moreover, GenAI's output can be very sensitive to the input parameters or settings \cite{liu_jailbreaking_2023,sun_automatic_2023}. Hallucination is another common concern with AI-generated content, as they can be imaginary or fictional \cite{alkaissi_artificial_nodate}. The promise of AI automation might be closer than ever in SE \cite{carleton_architecting_2022}, when these open challenges are sufficiently addressed.




Research agenda is a popular type of work for guiding and organizing research efforts in certain research areas \cite{jansen_sense_2009,sengupta_research_2006,bosch_engineering_2021,sriram_research_2010,france_model-driven_2007,papazoglou_service-oriented_2008}. It often includes a review of existing work, and a presentation of directions, visions and priorities, enabling researchers to make meaningful contributions to their respective fields and to address pressing challenges. This research agenda is driven by past and current work on GenAI for SE. Based on focus groups and a literature review, we portray the current research on GenAI in SE and present the research agenda with open challenges for future research. While we do not emphasize the comprehensiveness of the literature review due to the fast-moving nature of the topic, the focus groups present practical expectations of the future roles of GenAI in software development. The research agenda is organized into 11 areas of concern, which are: (1)~Requirements Engineering, (2)~Software Design, (3)~Sofware Implementation, (4)~Quality Assurance, (5)~Software Maintenance and Evolution, (6)~Software Processes and Tools, (7)~Engineering Management, (8)~Professional Competencies, (9)~Software Engineering Education, (10)~Macro Aspects and (11)~Fundamental concerns of GenAI. 


Two international events have provided a ground for this work. As a part of the first international workshop on AI-assisted Agile Software Development\footnote{https://www.agilealliance.org/xp2023/ai-assisted-agile/}, we ran our first focus group to explore the benefits and challenges of AI-assisted Agile software development. At the Requirements Engineering (RE) conference, we organized the second international workshop on Requirements Engineering for Software Startups and Emerging Technologies (RESET) \footnote{https://resetworkshop.org/} with a special theme on GenAI and Requirements Engineering. Consequently, the identified open challenges reflect not just the existing gaps in literature but also draw from the practical, value-driven insights of the co-authors.

The paper is structured as follows: we clarify the meaning of GenAI and what we know about GenAI in SE in Section 2, and we will present our research approach in Section 3. Section 4 presents our Research Agenda. Finally, Section 5 is Outlook and Conclusions.

\section{Background}

Application of AI/ML has a long history in SE research \cite{barenkamp2020applications,SE4AIMartinez2022,kotti_machine_2023}. The use of GenAI specifically, however, is a more emerging topic. While the promise of GenAI has been acknowledged for some time, progress in the research area has been rapid. Though some papers have already explored the application of GPT-2 to code generation \cite{paik2021improving}, for example, GenAI was not a prominent research area in SE until 2020. Following the recent improvements in the performance of these systems, especially the release of services such as GitHub Copilot and ChatGPT-3, research interest has now surged across disciplines, including SE. Numerous papers are currently available through self-archiving repositories such as arXiv\footnote{https://arxiv.org/}, and paperwithcode\footnote{https://paperswithcode.com/}. To prepare readers for further content in our research agenda, we present in this section relevant terms and definitions (Section 2.1), the historical development of GenAI (Section 2.2) and fundamentals on Large Language Models (Section 2.3)

\subsection{Terminologies}
Generative modelling is an AI technique that generates synthetic artifacts by analyzing training examples, learning their patterns and distribution, and then creating realistic facsimiles \cite{jovanovic_generative_2022}. GenAI uses generative modelling and advances in deep learning (DL) to produce diverse content at scale by utilizing existing media such as text, graphics, audio, and video. The following terms are relevant to GenAI:
 \begin{itemize}
     \item \emph{AI-Generated Content (AIGC)} is content created by AI algorithms without human intervention.
     \item \emph{Fine-Tuning (FT)} updates the weights of a pre-trained model by training on supervised labels specific to the desired task~\cite{brown_language_2020}.
     \item \emph{Few shot} training happens when an AI model is given a few demonstrations of the task at inference time as conditioning~\cite{brown_language_2020}.
     \item \emph{Generative Pre-trained Transformer (GPT)}: a machine learning model that uses unsupervised and supervised learning techniques to understand and generate human-like language~\cite{radford_improving_2018}.
     \item \emph{Natural Language Processing (NLP)} is a branch of AI that focuses on the interaction between computers and human language. NLP involves the development of algorithms and models that allow computers to understand, interpret, and generate human language.
     \item \emph{Language model} is a statistical (AI) model trained to predict the next word in a sequence and applied for several NLP tasks, i.e., classification and generation~\cite{lund_chatting_2023}.
     \item \emph{Large language model (LLM)} is a language model with a substantially large number of weights and parameters and a complex training architecture that can perform a variety of NLP tasks, including generating and classifying text, conversationally answering questions~\cite{brown_language_2020}. While the concept \textit{LLM} is currently widely used to describe a subset of GenAI models, especially out on the field and in the media, what exactly constitutes 'large' is unclear and the concept is in this regard vague. Nonetheless, given its widespread use, we utilize the concept in this paper as well, acknowledging this limitation.     
     \item \emph{Prompt} is an instruction or discussion topic a user provides for the AI model to respond to 
     \item \emph{Prompt Engineering} is the process of designing and refining prompts to instruct or query LLMs to achieve a desired outcome effectively.

 \end{itemize}

\subsection{History of GenAI}
To better understand the nature of GenAI and its foundation, we present a brief history of AI in the last 80 years: 

\begin{enumerate}
    \item Early Beginnings (1950s-1980s):
    Since 1950s, computer scientists had already explored the idea of creating computer programs that could generate human-like responses in natural language. Since 1960s, expert systems gained popularity. These systems used knowledge representation and rule-based reasoning to solve specific problems, demonstrating the potential of AI to generate intelligent outputs. Since 1970s, researchers began developing early NLP systems, focusing on tasks like machine translation, speech recognition, and text generation. Systems like ELIZA (1966) and SHRDLU (1970) showcased early attempts at natural language generation.
    \item Rule-based Systems and Neural Networks (1980s-1990s): Rule-based and expert systems continued to evolve during this period, with advancements in knowledge representation and inference engines. Neural networks, inspired by the structure of the human brain, gained attention in the 1980s. Researchers like Geoffrey Hinton and Yann LeCun made significant contributions to the development of neural networks, which are fundamental to GenAI.
    \item Rise of Machine Learning (1990s-2000s):
    Machine learning techniques, including decision trees, support vector machines, and Bayesian networks, started becoming more prevalent. These methods laid the groundwork for GenAI by improving pattern recognition and prediction.
    \item Deep Learning Resurgence (2010-2015):
    Deep learning, powered by advances in hardware and the availability of large datasets, experienced a resurgence in the 2010s. Convolutional Neural Networks (CNNs) and Recurrent Neural Networks (RNNs) emerged as powerful tools for generative tasks such as image generation and text generation. Generative Adversarial Networks (GANs), introduced by Ian Goodfellow and his colleagues in 2014, revolutionized GenAI. GANs introduced a new paradigm for training generative models by using a two-network adversarial framework. Reinforcement learning also made strides in GenAI, especially in game-related domains.
    \item Transformers and BERT (2015-now): Transformers, introduced in a seminal paper titled ``Attention Is All You Nee'' by Vaswani et al. in 2017, became the backbone of many state-of-the-art NLP models. Models like BERT (Bidirectional Encoder Representations from Transformers) showed remarkable progress in language understanding and generation tasks. GenAI has found applications in various domains, including natural language generation, image synthesis, music composition, and more. Chatbots, language models like GPT-3, and creative AI tools have become prominent examples of GenAI in software implementation.
\end{enumerate}

\subsection{Fundametals on LLMs}
LLMs are AI systems trained to understand and process natural language. They are a type of ML model that uses very deep artificial neural networks and can process vast amounts of language data. An LLM can be trained on a very large dataset comprising millions of sentences and words. The training process involves predicting the next word in a sentence based on the previous word, allowing the model to ``lear'' the grammar and syntax of that language. With powerful computing capabilities and a large number of parameters, this model can learn the complex relationships in language and produce natural-sounding sentences. Current LLMs have made significant progress in natural language processing tasks, including machine translation, headline generation, question answering, and automatic text generation. They can generate high-quality natural text, closely aligned with the content and context provided to them.

For an incomplete sentence, for example: ``The book is on the'', these models use training data to produce a probability distribution to determine the most likely next word, e.g., ``table'' or ``bookshelf''. Initial efforts to construct large-scale language models used N-gram methods and simple smoothing techniques \cite{brants2007large} \cite{heafield2013scalable}. More advanced methods used various neural network architectures, such as feedforward networks \cite{bengio2000neural} and recurrent networks \cite{mikolov2010recurrent}, for the language modelling task. This also led to the development of word embeddings and related techniques to map words to semantically meaningful representations \cite{mikolov2013efficient}. The Transformer architecture \cite{vaswani2017attention}, originally developed for machine translation, sparked new interest in language models, leading to the development of contextualized word embeddings \cite{lee2018pre} and Generative Pre-trained Transformers (GPTs) \cite{radford2019language}. Recently, a common approach to improving model performance has been to increase parameter size and training data. This has resulted in surprising leaps in the machine's ability to process natural language.

\vspace{.2cm}
\textbf{Parameter-Efficient Fine-Tuning}

In October 2018, the BERT Large model \cite{devlin2019bert} was trained with 350 million parameters, making it the largest Transformer model ever publicly disclosed up to that point. At the time, even the most advanced hardware struggled to fine-tune this model. The ``out of memory'' issue when using the large BERT model then became a significant obstacle to overcome. Five years later, new models were introduced with a staggering 540 trillion parameters \cite{chowdhery2022palm}, a more than 1500-fold increase. However, the RAM capacity of each GPU has increased less than tenfold (max 80Gb) due to the high cost of high-bandwidth memory. Model sizes are increasing much faster than computational resources, making fine-tuning large models for smaller tasks impractical for most users. 
In-context learning, the ability of an AI model to generate responses or make predictions based on the specific context provided to it \cite{radford2019language}, represents the ongoing advancements in the field of natural language processing. However, the context limitation of Transformers reduces the training dataset size to just a few examples. This limitation, combined with inconsistent performance, presents a new challenge. Furthermore, expanding context size significantly increases computational costs.

\begin{figure}[!htb]
        \centering
        \includegraphics[scale=0.48]{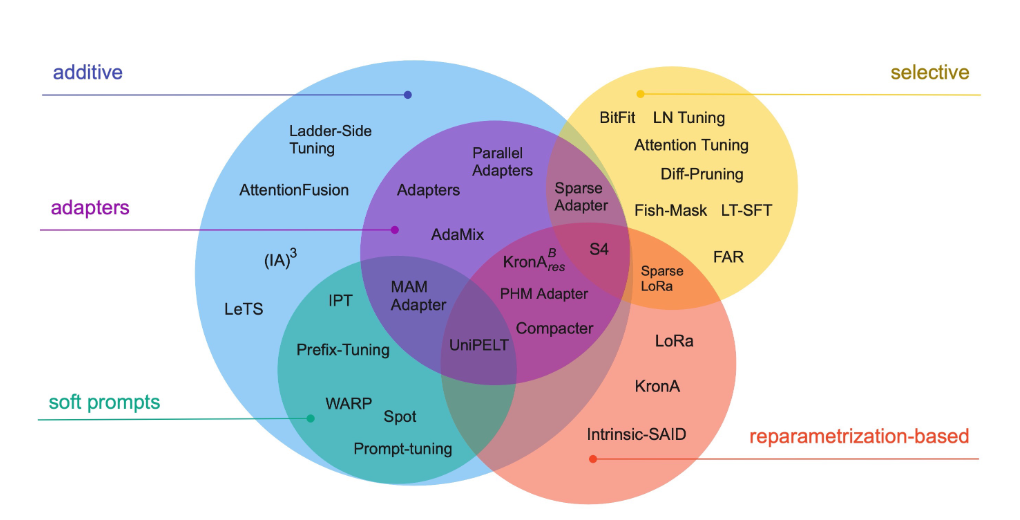}
        \caption[Several optimal parameter fine-tuning methods]{Several optimal parameter fine-tuning methods \cite{lialin2023scaling}}
        \label{fig:lialin2023scaling}
\end{figure}

Parameter-Efficient Fine-Tuning methods have been researched to address these challenges by training only a small subset of parameters, which might be a part of the existing model parameters or a new set of parameters added to the original model. These methods vary in terms of parameter efficiency, memory efficiency, training speed, model output quality, and additional computational overhead (if any).

\section{Research Approach}

As indicated earlier, this paper aims to present a research agenda listing open Research Questions regarding GenAI in SE.  We conducted a literature review (Section 3.1) and a series of focused groups to achieve this outcome (Section 3.2 and Section 3.3). The overall research process is presented in Figure \ref{fig:Rprocess}. 


\subsection{Literature review} 

A comprehensive and systematic review might not be suitable for research on GenAI in software development at the time this research being conducted. Firstly, research work is rapidly conducted and published on the topic, hence rendering the findings of a comprehensive review probably outdated shortly after publication. Secondly, we found a lot of relevant work as gray literature from non-traditional sources such as preprints, technical reports, and online forums. These sources may not be as rigorously reviewed or validated as peer-reviewed academic papers, making it difficult to assess their quality and reliability. Thirdly, we would like to publish the agenda as soon as possible to provide a reference for future research. A systematic literature review would consume extensive effort and time, which might be obsolete by the time the review is complete.

Our strategy is to conduct focused, periodic reviews to capture the most current and relevant information without the extensive resource commitment of a comprehensive review. This approach allows for agility in keeping up with the latest developments without claiming comprehensiveness and repeatability. Our search approach involves two channels:
 \begin{itemize}
     \item Online portals: we used Google Scholar and Scopus to search with a formulated search string
     \item Gray literature sources: we search for papers from Arxiv  and PaperwithCode
     \item Forward and backward snowballing: we scan citations forward and backward from the articles included in our review.
 \end{itemize}
 
 We used the search terms in Google Scholar and achieved the result (latest searched date October 2023). Google Scholar has the advantages of comprehensiveness and free access. It also covers grey literature where we found a significant number of research on GenAI at the searching time.

\subsection{Focus groups}
We conducted four structured working sections as focus groups to identify, refine, and prioritize Research Questions on GenAI for SE. Focus groups have been used as a means to gather data in SE research \cite{martakis_handling_2013,kontio_focus_2008,kontio_using_2004,dingsoyr_team_2013}
Focus group sessions produce mainly qualitative information about the objects of study. The benefits of focus groups are that they produce insightful information, and the method is fairly inexpensive and fast to perform. This differs from a brainstorming section where participants are selected and navigated by a moderator, following a structured protocol so that the discussions stay focused \cite{kontio_focus_2008,edmunds_focus_2000}. Kontio et al. also suggest online focus groups with a lot of advantages, such as group thinking, no travel cost, anonymous contribution, and support for large groups \cite{kontio_focus_2008}. The value of the method is sensitive to the experience and insight of participants. Hence, we presented the information of the participants in Table \ref{tab:demographic}.

\begin{table}[]
\caption{Demographic information of participants in workshops}
\label{tab:demographic}
\begin{tabular}{p{1.5cm}p{3.5cm}p{4.5cm}p{2cm}}
\hline
Id & Background                                   & Relevant Experience                                              & No of focus groups \\
\hline
P01            & Dr, Asst. Prof. in Software Engineering      & Adopt ChatGPT in Agile context                                   & 4                 \\
P02            & Prof. in Software Engineering and Applied AI & 3+ years research and teaching on AI and SE                      & 4                 \\
P03            & Dr in Software Engineering                   & Research about GenAI in Agile context                            & 4                 \\
P04            & Dr in Software Engineering                   & 5+ year research about NLP, applied AI in requirements engineering & 3                 \\
P05            & Dr in Software Engineering                   & Adopt ChatGPT in Agile context                                   & 4                 \\
P06            & Prof. in Software Engineering and Applied AI & 5+ years research and teaching on AI and SE                      & 2                 \\
P07            & Dr. in Software Engineering      & Adopt ChatGPT in Agile context                                                                  &          2         \\
P08            &  Asst. Prof. in Software Engineering                                            &  Pionner researchers in ChatGPT for Software Engineering                                                                & 2                  \\
P09            &   Dr in Software Engineering                   & Research and conducted Systematic Literature Review on GenAI for SE                                 & 4                    \\
P10            &     Dr. in Software Engineering      & Research on Applied AI   & 2                 \\
P11            &  Dr. in Software Engineering      & Adopt ChatGPT in teaching Software Engineering classes                                                                 &         4                  \\
P12            &    Prof. in Software Engineering and Applied AI & 5+ years research and teaching on AI and SE                      & 4                                    \\
P13            &       Prof. in Software Engineering and Applied AI & 10+ years research and teaching on AI/ML for SE                      & 2                  \\
P14            &    Dr. in Software Engineering, Certified Scrum Master                                          &   10+ years working in Agile projects, adopting ChatGPT in professional work                                                               &  4 \\         P15            &      Dr. in Software Engineering & Research and teaching Applied AI in SE       & 2       \\ 
\hline
\end{tabular}
\end{table}

\begin{figure}[t]
\includegraphics[width=0.5\linewidth]{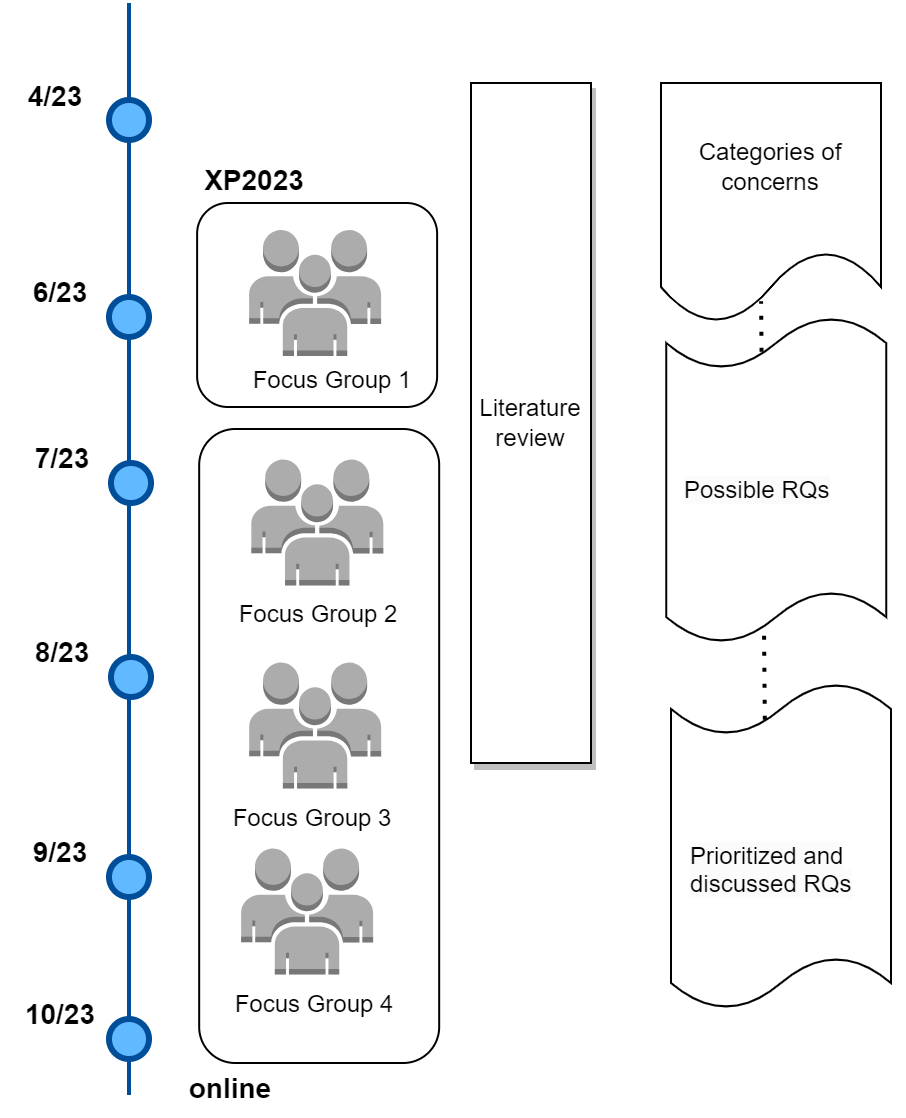}
\centering
\caption{Research Agenda on GenAI for Software Engineering}~\label{fig:Rprocess}
\end{figure}

Focus groups' timeline is from April 2023 to September 2023 (as shown in Figure \ref{fig:Rprocess}). All participants are SE researchers who have experience or interest in the topic, as shown in Table \ref{tab:demographic}. For each focus group, we developed a plan, which included the agenda of
the section and a set of exercises for the participants. Each focus group lasted between 2 to 3 hours. To capture data from focus groups, we use moderator's notes, Miro boards \footnote{https://miro.com}, and recording in case of online sessions. 

The focus of each focus group is different:
\begin{itemize}
    \item Focus group 1 (Exploratory brainstorming): Brainstorming was aimed at exploring ideas for potential opportunities and challenges when adopting GenAI in software development activities. We discussed a question, ``Which SE areas will benefit from GenAI tools, such as ChatGPT and Copilot?''. We used SWEBOK areas to initialize the conversations. At the end of the group, 11 categories were created. Each category is assigned a section leader who is a co-author responsible for synthesizing content for the section.
    \item Focus group 2 (Exploratory brainstorming):  We discussed ``What would be an interesting research topic for GenAI and Software Development and management?''. We initiated the discussion about ``What could be interesting Research Questions for an empirical study on GenAI in SE?'' Some questions were generated during the working session. 
    \item Focus group 3 (Validating brainstorming): prior to the meeting, a list of possible RQs was made. The list included 121 RQs. From this meeting, we aimed to validate the questions, if they are correctly and consistently interpreted, and if they are reasonable and meaningful to all participants of the meeting. Participants had sufficient time to review thoroughly and critically the RQs list. Each person left a comment for every question about their meaningfulness and feasibility. After this step, the RQs were revised and restructured.
    \item Focus group 4 (Validating brainstorming): The discussion was conducted in subgroups, each group focused on one particular SE category. Prior to this session, the question list was finalized. In total, there are 78 RQs in 11 categories. RQs that are consensusly not practically important or meaningful are excluded. We also discussed and ranked RQs according to their novelty and timeliness of each RQs. However, we did not achieve consensus and a complete list of rankings for all RQs. Therefore, we decided not to present the ranking in this work. 
\end{itemize}


\begin{figure}[t]
\includegraphics[width=\linewidth]{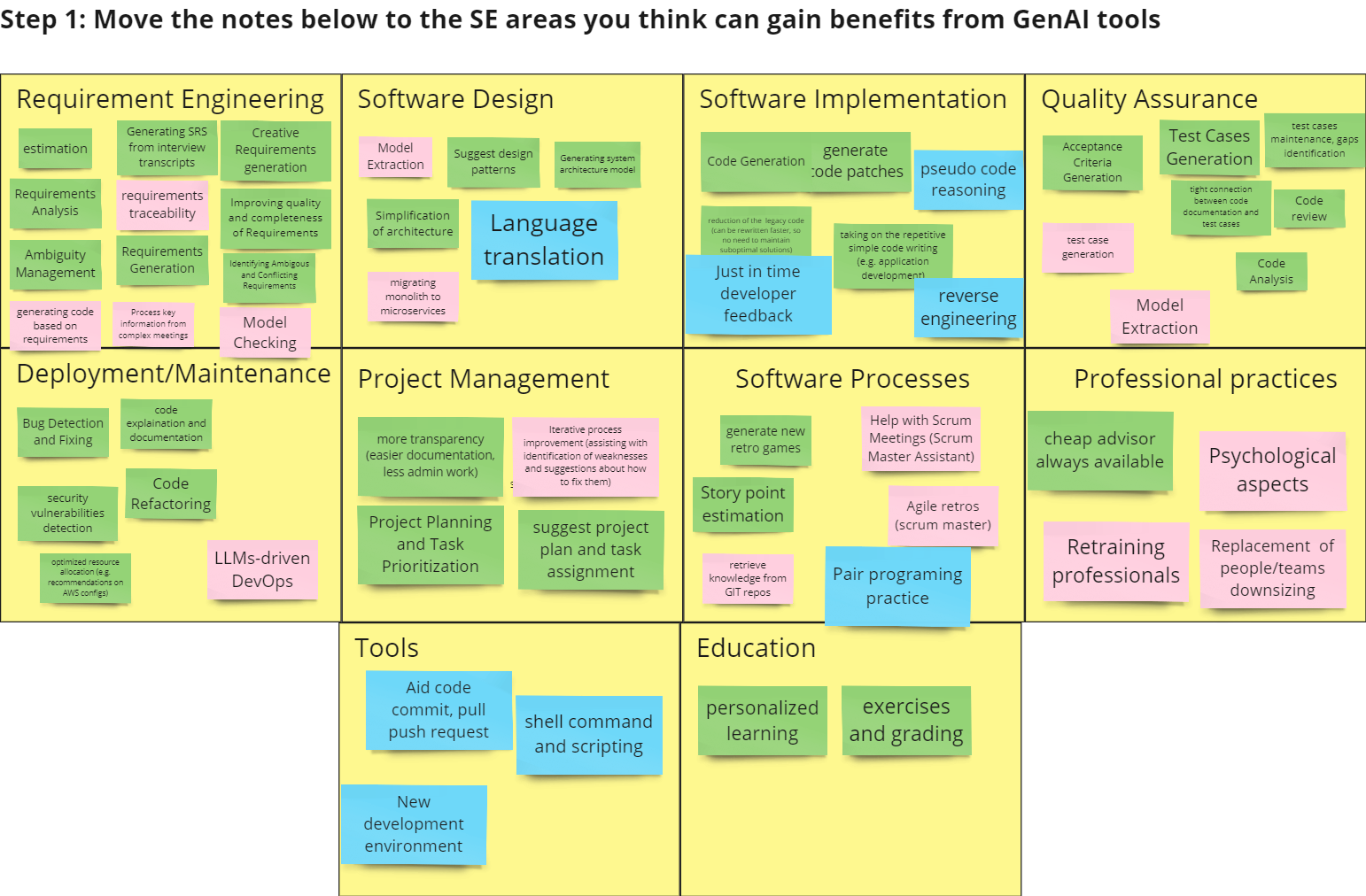}
\centering
\caption{An outcome from Focus Group 2}~\label{fig:workshop1}
\end{figure}




\subsection{Threats to validity}
According to Kontio et al. \cite{kontio_using_2004,kontio_focus_2008}threats of the focus groups method in SE studies include group dynamics, social
acceptability, hidden agenda, secrecy, and limited comprehension. To minimize the group dynamics threat, i.e., uncontrollable flow of discussion, all sections were coordinated by the main author, strictly following the pre-determined agenda with time control. Social acceptability can influence discussion results, i.e. participants contribute biased opinions because of positive or negative perceptions from others. We introduce clear rules for contribution at the beginning of focus groups. For online sessions, Miro is used so participants can anonymously provide their opinions. As time is relatively limited for an online focus group, some complex issues or points might not be not necessarily understood by all participants (limited comprehension). This threat is mitigated by organizing focus groups following up with each other. Between the sessions, participants were required to do their "homework" so discussing points could be studied or reflected further beyond the scope of the focus groups. Four focus groups occurring in four months give sufficient time for individual and group reflection on the topic.

\section{Research Agenda}

\begin{figure}[t]
\includegraphics[width=0.9\linewidth]{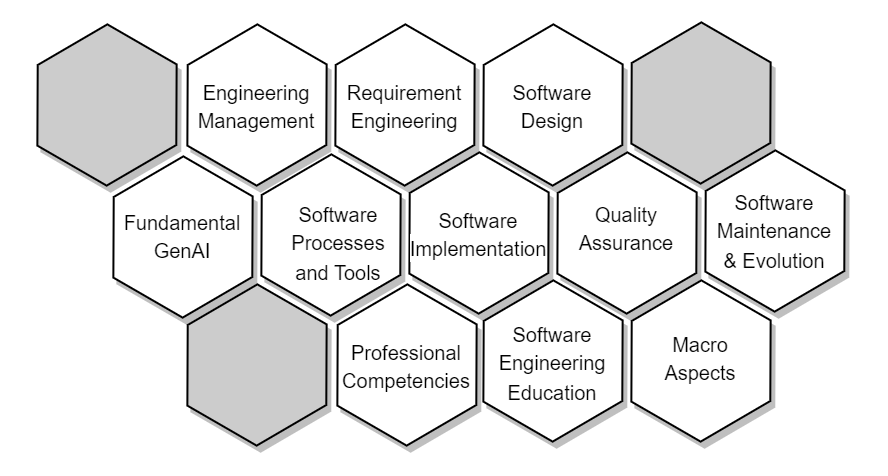}
\centering
\caption{Research Agenda on GenAI for Software Engineering}~\label{fig:Agenda}
\end{figure}

We grouped the research concerns and questions into eleven tracks based on the thematic similarities and differences of the questions. While this grouping is one of the several possible ways to create the categories, it served the purpose of easing the presentation and discussion of the research agenda, shown in Figure \ref{fig:Agenda}. For each category, we present (1) its historical context, (2) key RQs, (3) State-of-the-art, (4) current challenges and limitations, and (6) future prospects.

\subsection{GenAI in Requirements Engineering}

\subsubsection{Historical Context.}
Requirements Engineering (RE) plays a pivotal role in the success of software projects. Although numerous techniques exist to support RE activities, specifying the right requirements has remained challenging as requirements are seldom ready for collection in the initial stages of development~\cite{Sosnowski_2021}. Stakeholders rarely know what they really need, typically having only a vague understanding of their needs at the beginning of the project. These needs evolve during the project largely due to the dynamism inherent in business and software development environments~\cite{Przybylek_2018}. Among Agile projects, RE has new challenges, such as minimal documentation, customer availability, lack of well-understood RE practices, neglecting non-functional requirements, lack of the big picture, inaccurate effort estimation and lack of automated support~\cite{Inayat_2015,ramesh2010agile}. Moreover, several other avenues in RE have remained relatively under-explored or under-developed, even in agile. These include, among others, the traceability and maintenance of requirements, stringent validation and verification processes, coverage of human factors in RE, ethical considerations while employing automation to RE or RE for AI software systems~\cite{ahmad2023requirementsFramework}, coverage of domain specificity in requirements (e.g., stringent regulatory compliance for healthcare domain). RE, replete with its historical challenges and contemporary complexities, will need to adapt and evolve its role to the landscape of modern software development with technologies like GenAI, which alleviates some of these challenges~\cite{ahmad2023requirements}.

\subsubsection{Key RQs.}
Regardless of the order in which RE tasks are performed or their application in either a waterfall or agile framework, the foundational RE stages and corresponding tasks remain consistent. This means that one still needs to identify stakeholders, elicit requirements, specify them in a format of choice, e.g., as user stories or shall-style requirements, analyze them, and perform other V\&V tasks using them. Based on these considerations, we identified the following RQs.
\begin{enumerate}
    \item How can GenAI support requirements elicitation?
    \item How can GenAI effectively generate requirements specifications from high-level user inputs?
    \item How can GenAI facilitate the automatic validation of requirements against domain-specific constraints and regulations?
    \item How can GenAI be used to predict change requests?
    \item What are the challenges and threats of adopting GenAI for RE tasks in different RE stages (pre-elicitation, elicitation, specification, analysis)?
\end{enumerate}


\subsubsection{State-of-the-art}
Many research endeavors seek to utilize GenAI in order to automate tasks such as requirements elicitation, analysis, and classification. The current literature is dominant by preliminary empirical evidence of the significant role that LLMs in RE \cite{white_chatgpt_2023,ronanki2023investigating,zhang2023adaptive,arulmohan2023extracting,ezzini2022automated,Moharil_2023,arora2023advancing}. For instance, White et al.~\cite{white_chatgpt_2023} introduced a catalog of prompt patterns, enabling stakeholders to assess the completeness and accuracy of software requirements interactively. Ronanki et al.~\cite{ronanki2023investigating} further substantiated this by comparing requirements generated by ChatGPT with those formulated by RE experts from both academia and industry. The findings indicate that LLM-generated requirements tend to be abstract yet consistent and understandable.

In addition to requirements elicitation, there's also a focus on requirements analysis and classification. Zhang et al.~\cite{zhang2023preliminary} developed a framework to empirically evaluate ChatGPT's ability in retrieving requirements information, specifically Non-Functional Requirements (NFR), features, and domain terms. Their evaluations highlight ChatGPT's capability in general requirements retrieval, despite certain limitations in specificity. In agile RE practice, Arulmohan et al.~\cite{arulmohan2023extracting} conducted experiments on tools like Visual Narrator and GPT-3.5, along with a conditional random field (CRF)-based approach, to automate domain concept extraction from agile product backlogs. Their work provides evidence of the advancements in automating domain concept extraction.  
Both Ezzini et al. [2022] \cite{ezzini2022automated} and Moharil et al. [2023] \cite{Moharil_2023} have tackled issues pertaining to anaphoric ambiguity in software requirements. Anaphoric ambiguity emerges when a pronoun can plausibly reference different entities, resulting in diverse interpretations among distinct readers \cite{ezzini2022automated}. Ezzini et al. [2022] \cite{ezzini2022automated} have developed an automated approach that tackles anaphora interpretation by leveraging SpanBERT. Conversely, Moharil et al. [2023] \cite{Moharil_2023} have introduced a tool based on BERT for the detection and identification of ambiguities. This tool can be utilized to generate a report for each target noun, offering details regarding the various contexts in which this noun has been employed within the project. Additionally, De Vito et al.~\cite{de2023echo} proposed the ECHO framework, designed to refine use case quality interactively with ChatGPT, drawing from practitioner feedback. The findings indicate the efficacy of the approach for identifying missing requirements or ambiguities and streamlining manual verification efforts. BERT-based models have also been applied in RE to improve the understanding of requirements and regulatory compliance via automated question-answering~\cite{abualhaija2022automated,Ezzini:23}.

Hey et al. [2020] \cite{Hey_2020} introduced NoRBERT, which involves fine-tuning BERT for various tasks in the domain of requirements classification. Their findings indicate that NoRBERT outperforms state-of-the-art approaches in most tasks. Luo et al. [2022] \cite{Luo_2022}, in a subsequent study, put forward a prompt-based learning approach for requirement classification using a BERT-based pre-trained language model (PRCBERT). PRCBERT utilizes flexible prompt templates to achieve accurate requirements classification. The conducted evaluation suggests that PRCBERT exhibits moderately superior classification performance compared to NoRBERT. Additionally, Chen et al.~\cite{chen2023use} evaluated GPT-4's capabilities in goal modeling with different abstraction levels of information, revealing its retained knowledge and ability to generate comprehensive goal models.

\subsubsection{Challenges and Limitations}
Automation of RE tasks with GenAI techniques is undoubtedly promising; however, it is paramount for practitioners to be aware of several risks to ensure a judicious blend of domain and RE expertise and AI's capabilities. Hallucination is a common challenge for adopting GenAI for all RE activities. GenAI techniques might generate requirements that appear sound but are either superfluous, incorrect, or inconsistent in a given domain context. Such inconsistencies can inadvertently lead to projects straying off course if not meticulously reviewed by stakeholders with relevant experience. Furthermore, some RE tasks (e.g., traceability) would require fine-tuning LLMs to the task data, but historically, not much RE data is publicly available to fine-tune these models accurately. We present some challenges and limitations of GenAI models regarding each RE stage, as below: 
\textbf{Pre-Elicitation Stage:} While LLMs can use preliminary research materials such as existing systems, universal documents, and app reviews to automate domain and stakeholder analysis, they sometimes fail to capture the broader context. Inaccurate identification of key domain concepts or overlooking intricate relationships can lead to suboptimal domain models.

\textbf{Elicitation Stage:} LLMs can process interview transcripts, survey responses, and observation notes to aid in the requirements elicitation process. However, the inherent biases of LLMs can affect the coding, analysis, and summarization of this data, potentially leading to skewed requirements.

\textbf{Specification Stage:} LLMs offer automated generation and classification of requirements. However, these generated requirements can sometimes lack clarity or coherence without proper human oversight. The reliance on LLMs for regulatory compliance or requirements prioritization might also result in overlooking important nuances in legal or organizational guidelines.

\textbf{Analysis Stage:} Using LLMs for requirements quality analysis, traceability, and change management can automate and streamline the RE process. However, LLMs might sometimes struggle with intricate defect detection, traceability challenges, or handling complex change scenarios, leading to gaps in the analysis.

\textbf{Validation Stage:} Generative models often lose the broader context, especially with intricate requirements, potentially leading to discrepancies or conflicts within the generated requirements. The onus then lies on SE teams to validate such AI-generated requirements or other automated outputs, introducing an additional layer of complexity and potential validation challenges. 

Overall, the ethical and security ramifications of delegating pivotal RE tasks to AI also warrant attention, mainly in safety-critical systems or application domains such as healthcare, defense, or cybersecurity. Lastly, a challenge that spans beyond just RE is the potential bias and fairness issues. If unchecked, generative models can perpetuate or even intensify biases in their foundational data. This can lead to skewed requirements that might not represent the diverse needs of end-users~\cite{hidellaarachchi2023influence}.

\subsubsection{Future Prospects.}
The future holds intriguing prospects regarding automating most RE tasks and making requirements engineers' lives easier. We believe that the future way of working will be AI-human collaborative platforms where GenAI assists domain experts, requirements engineers, and users in real-time, allowing for instantaneous feedback loops and iterative refining of requirements might be the new norm~\cite{li_tackling_2023}. Such platforms could also offer visualization tools that help stakeholders understand how the AI interprets given inputs, fostering a more transparent understanding of system requirements and the generated outputs. We expect to witness LLMs capable of deeper contextual understanding, reducing inaccuracies in the pre-elicitation and elicitation stages. An important prospect would be allowing fine-tuning of models without exposing sensitive information from requirements (i.e., data leakage). Considering the ethical and safety concerns, there is an evident need for establishing robust frameworks and guidelines for the responsible use of AI in RE. In the near future, the research could be directed at creating models with built-in ethical considerations capable of addressing the known biases, ensuring that the generated requirements uphold the standards and are truly representative and inclusive.

\subsection{GenAI in Software Design}


\subsubsection{Historical Context}
Design and architecture decisions are at the core of every software system. While some well-known structures are easy to replicate and create templates, some decisions need an evaluation of the trade-offs involved, especially on how these decisions affect the system's quality attributes. In this context, choosing the pattern or solution is just the last step of the process, which requires identifying all factors that influence and might be affected by the decision. We can find in the literature several works that proposed solutions to automate design decisions, mainly based on the choice of patterns, such as based on questions and answers \cite{Issaoui2015}, based on anti-patterns detection \cite{Nahar2016improved}, based on text classification \cite{Van2020}, and based on ontologies \cite{Bou2018}. However, it is a fact that these approaches are not yet widely used in industry, and GenAI is attracting interest from professionals as a promising approach to be adopted for design decisions in real projects. 

\subsubsection{Key RQs}
Similar to Requirements Engineering, it is necessary to explore the possible adoption of GenAI in all software design activities. As listed below, we present RQs addressing specific software design activities to illustrate open challenges in GenAI and software design:

\begin{enumerate}
    \item How can GenAI assist in identifying and selecting appropriate design and architectural patterns based on requirements and constraints?
    \item What strategies can be adopted to promote collaboration for decision-making between professionals and GenAI in the software design process?
    \item What are the limitations and risks associated with using GenAI in software design, and how can these be mitigated to ensure their reliability and validity?
    \item How can GenAI be employed in a continuous software design process, automating the identification of improvements and refactoring to optimize key quality attributes?
    \item How can generative AI be utilized to automate the design and implementation of user interfaces, improving user experience in software applications?
    \item How can GenAI optimize the trade-offs between different quality attributes, such as performance, scalability, security, and maintainability?
    \item What strategies can be employed to enable GenAI to adapt and evolve system architectures over time, considering changing requirements, technologies, and business contexts?
\end{enumerate}


\subsubsection{State-of-the-art}
We do not find many papers exploring GenAI or LLM for software design activities. Ahmad et al. described a case study of collaboration between a novice software architect and ChatGPT to architect a service-based
software \cite{chatgptarch2023}. Herold et al. proposed a conceptual framework for applying machine learning to mitigate architecture degradation \cite{herold_towards_2020}. A recent exploratory study evaluated how well ChatGPT can answer that describes a context and answer which design pattern should be applied \cite{Jota2023}, demonstrating the potential of such a tool as a valuable resource to help developers. However, in the cases where ChatGPT give a wrong answer, a qualitative evaluation pointed out that the answer could mislead the developers. Stojanović et al. presented a small experiment where ChatGPT identified microservices and their dependencies from three pieces of system descriptions \cite{stojanovic_application_2023}. Feldt et al. proposed a hierarchy of design architecture for GenAI-based software testing agents \cite{feldt_towards_2023}.

\subsubsection{Challenges and Limitations}
The challenges of GenAI in software design go beyond just receiving the correct answer for a design problem. It is essential to understand what role it should play and how it will interact with the software developers and architects in a team. Considering that design decisions happen at several moments during a project, it is also important to understand how a GenAI can be integrated into that process. 

A limitation of the current literature on the topic is the lack of studies in real and industrial environments. While the techniques were experimented with in a controlled environment, it is still unknown how useful they can really be in practice. While studies show that the usage of AI can provide correct answers to software design problems with good accuracy, inside a real project environment, it is still a challenge to understand which information needs to be provided and if the answer can really be insightful and not just bring information that the team already knows. While the recent works delimit the scope in a limited set of patterns, usually the classic Gang of Four (Gof) patterns \cite{gamma1995elements}, for industrial usage, answers are expected to consider a wide spectrum of solutions. Considering the findings of a recent study that exposed several problems in how non-functional requirements are handled \cite{viviani2023empirical}, real projects might not have all the information needed as input for a GenAI tool to give an accurate answer. 

\subsubsection{Future Prospects}
With access to tools that can give useful answers, the key challenge for the future of GenAI in software design is how GenAI tools will fit into the existing processes to enable its usage more consistently. Considering that the design of the architecture happens continuously, using these tools only as an oracle to talk about the solutions as a personal initiative might be an underutilization of its potential. To address the RQs in this domain, it is imperative to comprehend the roles GenAI tools can assume in the software design process and the approaches for engaging with them using prompt patterns, as discussed in \cite{white_prompt_2023}. It is expected that new good and bad software design practices will soon emerge from the first experiences with this kind of tool.

\subsection{GenAI in Software Implementation}


\subsubsection{Historical Context}
Software implementation is the crucial phase where the designed solution is transformed into a functional working application. With the advent of object-oriented programming in the 1990s, software systems have rapidly grown in size and complexity. Automation is an important research direction to increase productivity while ensuring the quality of software code. Modern Integrated Development Environments (IDEs) with features, i.e. compiling, deploying, debugging and even code generating, are continuously evolving to maximize the support for software developers.

However, implementing software systems in industries today presents several challenges. First, there is the constant pressure to keep up with rapidly evolving technology, which often leads to compatibility issues with legacy systems. Second, ensuring security and data privacy has become increasingly complex as cyber threats continue to evolve. Third, scalability and performance optimization are critical as software systems must handle growing amounts of data and users. Lastly, the shortage of skilled software engineers and the rising cost of software development further exacerbate these challenges. Thus, successful software implementation requires careful planning, ongoing maintenance, and adaptability to stay ahead in today's competitive technological landscape.

\begin{figure}[!htb]
        \centering
        \includegraphics[scale=0.33]{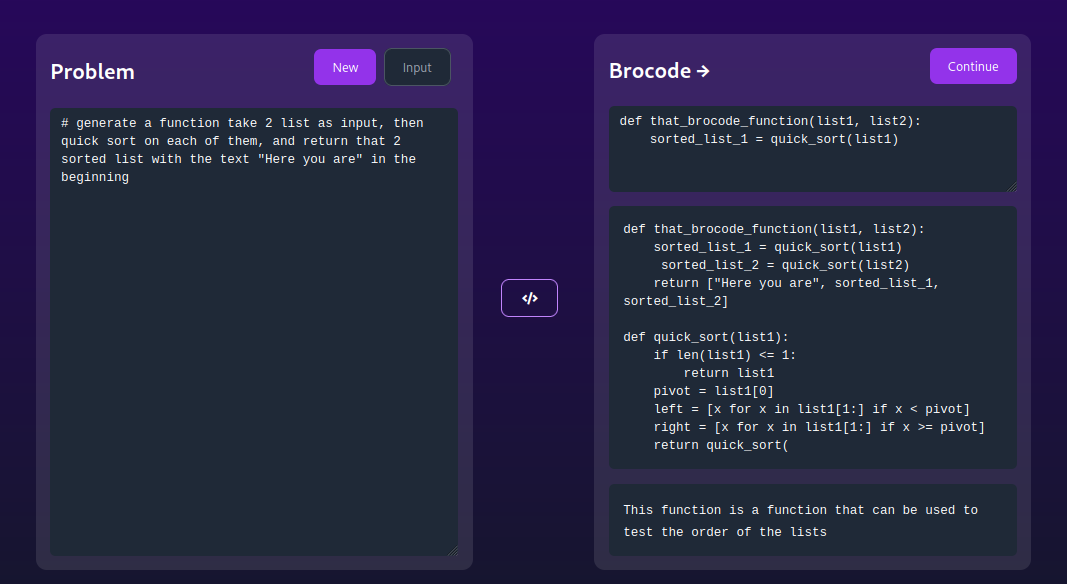}
        \caption[Results of the proposed model for simple logic requests]{Results of the proposed model for simple logic requests}
        \label{fig:swe_ex_1}
\end{figure}

\begin{figure}[!htb]
        \centering
        \includegraphics[scale=0.3]{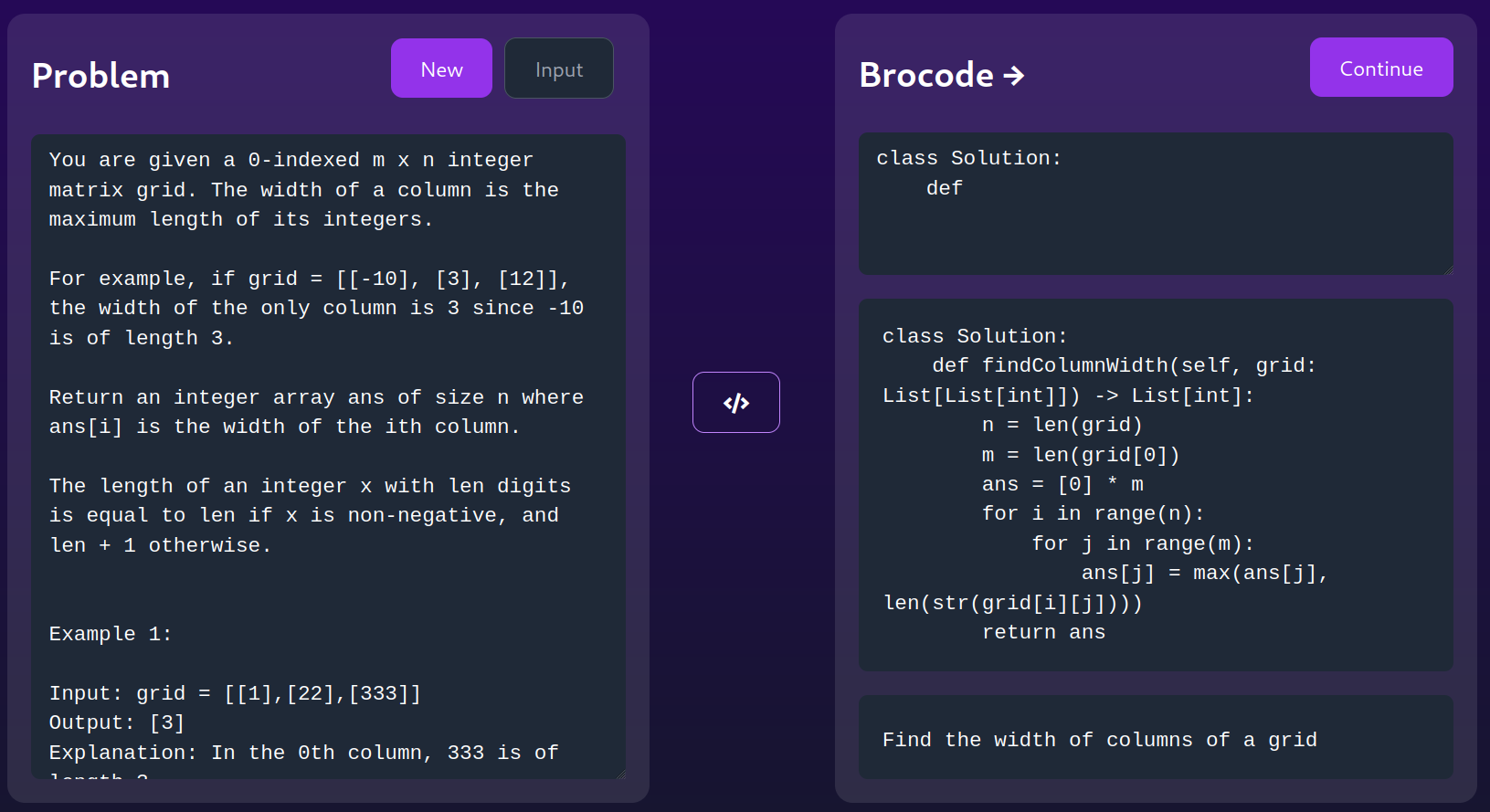}
        \caption[Results of the proposed model for complex requests]{Results of the proposed model for complex requests}
         \label{fig:swe_ex_2}
\end{figure}

\textit{Code generation} stands as a crucial challenge in both the field of GenAI and the software development industry. Its objective is to automatically generate code segments using requirements and descriptions expressed in natural language. Addressing this challenge can yield a multitude of advantages, including streamlining the programming process, improving the efficiency and accuracy of software development, reducing the time and effort required for coding, and ensuring the uniformity and correctness of the generated code. To showcase the capability of code generation, we requested ChatGPT with a simple task:"
Create a function that takes two arrays as input, then performs the quick sort algorithm on each array and returns the results of the two sorted arrays with the phrase 'Here you go'." (Figure \ref{fig:swe_ex_1}) To complex requests, such as a new algorithm problem taken from Leetcode. (Figure \ref{fig:swe_ex_2}). The result is a complete solution that, when submitted, is correct for all test cases on Leetcode\footnote{https://leetcode.com/}.

\subsubsection{Key RQs}
Considering the challenges identified for modeling and training GenAI architecture in the field of Software Implementation, We highlight particular RQs (RQs) related to the process of building a model that can be used in a real-world industrial environment

\begin{enumerate}
    \item To what extent GenAI-assisted programming, i.e. code summarization, code generation, code completion, comment generation, code search, and API search can be effectively integrated into practical software development projects? 
    \item 
    How can GenAI models be specialized for specific software domains, such as web development, game development, or embedded systems, to generate domain-specific code more effectively?
    \item How to ensure the correctness and reliability of generated results, carefully consider the potential for hidden malicious output that we are unaware of?
    \item 
    How can software companies both leverage the capacity of LLMs and secure their private data in software development?
    \item Does the outcome generated by a GenAI model have a significant bias towards the knowledge learned from LLMs while overlooking important input segments from private datasets of an enterprise?    \item How can the system effectively specify user intent when users communicate by progressively providing specifications in natural language
    \item  How can GenAI model be built, trained and retrained for a cost-based performance in various types of software implementation tasks?  \cite{hoffmann2022training} \cite{touvron2023llama}
    \item What needs to be done to achieve a Natural Language Interface for Coding where non-IT people can also interact and develop their wanted software?    
    \item What methods can be employed to validate AI-generated code against functional and non-functional requirements?
    \item  How can we ensure GenAI models comply with certain legal and regulation constraints \cite{stability-inc} \cite{github-inc} ?

\end{enumerate}


\subsubsection{State-of-the-art}


Developing code in an efficient, effective, and productive manner has always been an important topic for SE research. Since 2020, there has been extensive research about LLMs for various coding tasks, such as code generation, code completion, code summarization, code search, and comment generation. Various studies investigate the code generation performance of GitHub Copilot in different work contexts and across different Copilot versions, e.g.,
\cite{moradi_dakhel_github_2023,imai_is_2022,siddiq_zero-shot_2023}. Jiang et al. proposed a two-step pipeline that use input prompts to generate intermediate code, and then to debug this code \cite{jiang_selfevolve_2023}. The approach is promising to bring consistent efficacy improvement. Dong et al. treated LLMs as agents, letting multiple LLMs play distinct roles in addressing code generation tasks collaboratively and interactively \cite{Dong_2023}. Borji et al. presented a rigorous, categorized and systematic analysis of LLM code generation failures for ChatGPT \cite{borji_categorical_2023}. Eleven categories of failures, including reasoning, factual errors, mathematics, coding, and bias, are presented and discussed in their work. Sun et al. focus on users’ explainability needs for GenAI in three software engineering use cases: code generation based on natural language description (with Copilot), translation between different programming languages (with Transcoder), and code autocompletion (with Copilot) \cite{sun_investigating_2022}.

Several studies attempt to compare different code generation tools. Li et al. proposed an approach called AceCoder that surpasses existing LLM-based code generation tools, i.e. CodeX, CodeGeeX, CodeGen, and InCoder on several benchmarks \cite{Li_2023}. Doderlein et al. experimented with various input parameters for Copilot and Codex, finding that varying the input parameters can significantly improve the performance of LLMs in solving programming problems \cite{doderlein_piloting_2023}. Yetistiren et al. [74] presented a comprehensive evaluation of the performance of Copilot, CodeWhisperer, and ChatGPT, covering different aspects, including code validity, code correctness, code security, and code reliability. Their results show a wide degree of divergence in performance, motivating the need for further research and investigation \cite{yetistiren_evaluating_2023}

Salza et al. presented a novel LLM-driven code search approach by pre-training a BERT-based model on combinations of natural language and source code data and fine-tuning with data from StackOverflow \cite{salza_effectiveness_2023}. Chen et al. explored the setting of pre-train language models on code summarization and code search on different program languages \cite{chen_transferability_2022}. Guo et al. developed a model named CodeBERT with Transformer-based neural architecture for code search and code generation \cite{feng_codebert_2020}. The model has been widely popular and serves as a baseline for a lot of LLM-driven code generation approaches. For instance, Guo et al. proposed a model called GraphCodeBERT, a pre-trained model for programming language that considers the inherent structure of code. GraphCodeBert uses data flow in the pre-training stage and is texted on code search and code generation \cite{guo_graphcodebert_2021}.
Code retrieval, a practice to reuse existing code snippets in open-source repositories \cite{gu_cradle_2021}, is also experimented GenAI. Li et al. proposed a Generation-Augmented Code Retrieval framework for code retrieval task \cite{li_generation-augmented_2022}. 

Automatic API recommendation is also a sub-area of LLM applications. Wei et al. proposed CLEAR, an API recommendation approach that leverages BERT sentence embedding and contrastive learning \cite{wei_clear_2022}. Zhang et al. [387] developed ToolCoder, which combines API search tools with existing models to aid in code generation and API selection \cite{zhang_toolcoder_2023}. Patil et al. \cite{patil_gorilla_2023} developed a tool called Gorilla. This is a fine-tuned LLaMA-based model that is able to interact and perform with API, i.e. API call and verification. 

Comment generation is also an active research area, as source code is not always documented, and code and comments do not always co-evolve- \cite{Mastropaolo_2021}. Mastropaoplo et al. studied comment completion by validating the usage of Text-To-Text Transfer Transformer (T5) architecture in autocompleting a code comment \cite{Mastropaolo_2021}. Geng et al. adopted LLMs to generate comments that can fulfill developers' diverse intents \cite{geng_large_2023}. The authors showed that LLM could significantly outperform other supervised learning approaches in generating comments with multiple intents.

Several works have been done in an industrial setting. Vaithilingam et al. \cite{vaithilingam_expectation_2022} explored how LLMs (GitHub Copilot) are utilized in practice in programming, finding that even though they did not necessarily increase work performance, programmers still preferred using them. Barke et al. \cite{barke_grounded_2023}, in a similar study of how programmers utilize LLMs, highlighted two use modes: acceleration (next action is known and LLM is used to code faster) and exploration (next action is unknown and LLM is used to explore options). Denny et al. showed that Copilot could help solve 60 percent of programming problems and a potential way to learn writing code \cite{denny_conversing_2022}. Ziegler et al. \cite{ziegler_productivity_2022} explored potential measures of productivity of LLM use in programming, finding that the "rate with which shown suggestions are accepted" is a metric programmers themselves use to gauge productivity when utilizing LLMs.

Ouyang et al. empirically studied the nondeterminism of ChatGPT in code generation and found that over 60\% of the coding tasks have zero equal test output across different requests \cite{ouyang_llm_2023}. Pearce et al. discussed potential security issues related to code generated with GitHub Copilot \cite{pearce_asleep_2022}. Finally, in terms of programming practices, the use of LLM-based assistants such as GitHub copilot is often likened to pair programming  \cite{bird_taking_2023,imai_is_2022}. 

\subsubsection{Relevant technologies}


\begin{figure}[!htb]
\centering
\includegraphics[scale=0.23]{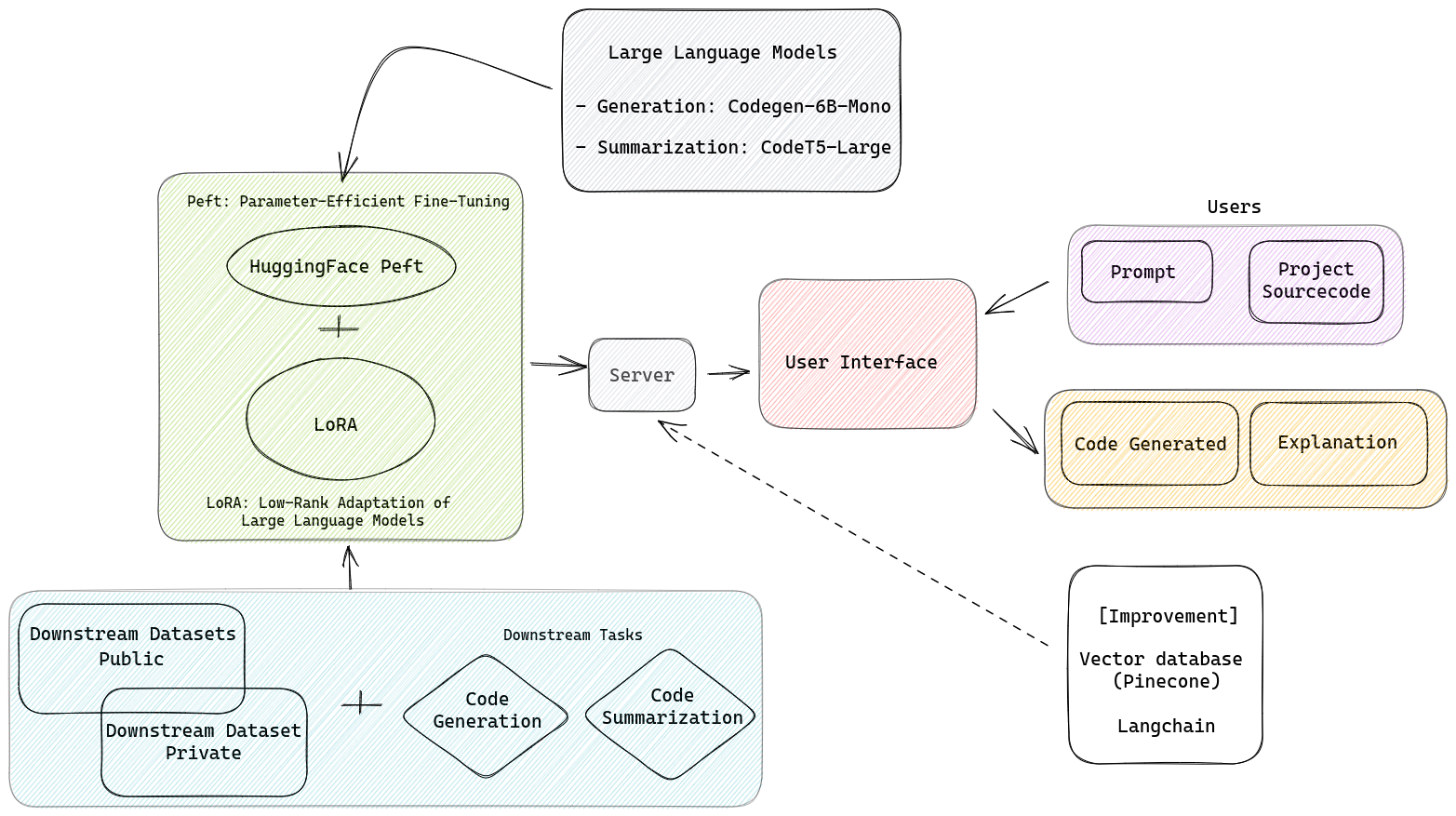}
\caption[Overview of the GenAI for code]{Overview of the GenAI for code}
\label{fig:architecture-overview}
\end{figure}

The flow of constructing the code generation model is depicted in Figure \ref{fig:architecture-overview}. As seen, the flow of the automatic code generation system is defined as follows:

\begin{table}
    \centering
    \begin{tabular}{|p{7cm}|p{6cm}|}
     \hline   Latest LLMs for Code Generation & Latest Parameter Fine-Tuning Methods \\
            \begin{itemize}
    \item CodeGen-Multi: \emph{SalesForce}            
    \item CodeT5+: \emph{SalesForce}
    \item SantaCoder: \emph{HuggingFace}
    \item StarCoder: \emph{HuggingFace}
    \item Code Llama: \emph{Facebook}
    \item PaLM-Coder: \emph{Google}
    \end{itemize}
    & 
    \begin{itemize}
    \item LoRA \cite{hu2021lora}
    \item AdaLoRA \cite{zhang2023adaptive}
    \item Prefix Tuning \cite{li-liang-2021-prefix} \cite{liu2022ptuning}
    \item P-Tuning \cite{liu2021gpt}
    \item Prompt Tunning \cite{lester2021power}
    \item MultiTask Prompt Tuning \cite{wang2023multitask}
    \item $(IA)^3$ \cite{liu2022fewshot}
    \end{itemize}        
        \\\hline
    \end{tabular}
    \caption{Latest development in LLM training and fine-tuning}
    \label{tab:my_label}
\end{table}


\begin{itemize}
\item[\textbf{Step 1:}] Research and utilize LLMs that have the best performance and an appropriate number of parameters for tasks related to automatic code generation and code summarization. In this case, the author employs two models: \textit{CodeGen-6B-Mono} for code generation (Generation) and \textit{CodeT5-Large} for code summarization.
\item[\textbf{Step 2:}] Implement methods to optimize parameter fine-tuning on \textit{specific private datasets} or carefully select public datasets that were not part of the training set of the initial large models (downstream datasets). Here, the author uses the \textit{LoRA method} and the HuggingFace Peft library\footnotemark[5] to handle the two defined downstream tasks.
\item[\textbf{Step 3:}] Store the trained model and develop a \textit{user interface}, enabling users to interact and experience automatic code generation and summarization based on the initial specific dataset.
\item[\textbf{Step 4:}] Develop secure storage mechanisms for user-specific projects and continue training on new user datasets.
\end{itemize}

\subsubsection{Challenges and Limitations}
Challenges and Limitations of current GenAI applications on software implementation can be presented under three aspects: (1) expressiveness, (2) program complexity, and (3) security and specification requirements.

\textbf{Heterogeneous inputs}
For an AI generation system to function effectively, the first unavoidable challenge is understanding the input requirements. The system needs to analyze and grasp specific requirements, determining what it needs to generate to satisfy those particular demands. Users' intentions can be expressed in heterogeneous manners, i.e. specifying logical rules, using input-output examples, or describing in natural language. Addressing all logical rules comprehensively is demanding, while input-output examples often don't sufficiently describe the entire problem. It can be argued that the ability to recognize and articulate user intentions is the key to the success of automated code generation systems, a challenge that traditional approaches have yet to solve.

\vspace{.2cm}
\textbf{Model Size}
With the increasing size and complexity of codebases and the emergence of numerous new technologies, understanding and generating code is also becoming more and more complex. GenAI model might need to include more and more dimensions, making it impractical and unrealistic for most people to fine-tune large models for smaller tasks. It's turning into a race dominated by big global tech corporations. Notable applications such as ChatGPT, Microsoft's Github Copilot], Google's Bard, and Amazon's Code Whisperer share a common goal: harnessing AI to make programming more accessible and efficient. This race is intensifying. In May 2023, HuggingFace launched its LLM BigCode \cite{li2023starcoder}, while Google announced its enhanced PaLM-2 model. Both made astonishing advancements, competing directly with Microsoft's famous GPT models. Recently, in August 2023, Facebook announced Code LLaMa, which is expected to give the open-source community easier access to this exciting race.

\vspace{.2cm}
\textbf{Security and Specification Requirements}
Security is also a major challenge for GenAI systems. Due to their vast and comprehensive knowledge base, they are in a position where they "know everything, but not specifically what." Alongside this, there's an increasing need for code security and very specific logical inputs in real-world scenarios. Consider not retraining the GenAI system with the code and knowledge of a specific (private) project but instead applying general knowledge to everyday tasks in the Software Industry. It can be easily predicted that the system won't be able to generate code satisfying complex real-world requirements. Applying the specific logic of each company's security project requires the system to grasp not just general knowledge but also the business and specific logic of each particular project. However, suppose an organization or individual tries to retrain a LLM on their project's source code. This is not very feasible. First, training an LLM requires robust hardware infrastructure that most individuals or organizations cannot guarantee. Secondly, the training time is usually significant. Lastly, the result of training is a massive weight set, consuming vast storage space and operational resources. Resources and efforts spent on training multiple different projects can be wastefully redundant, especially when users only need the model to work well on a few specific, smaller projects. On the contrary, focusing on training specific projects from the outset means we can't leverage the general knowledge LLMs offer, leading to lower-quality generated code. Thus, the author believes that while automated code generation might work well for general requirements using \textit{LLM}, there remain many challenges when applying them to specific and unique projects. To address these challenges, recent years have seen numerous studies dedicated to discovering methods to \textit{fine-tune LLM} using parameter optimization methods. Most of these approaches share a goal: applying general knowledge to specific problems under hardware constraints while retaining comparable efficiency and accuracy.


\subsubsection{Future Prospects}


The intrinsic appeal of GenAI on a global scale is undeniable. Ongoing research endeavors are significantly focused on enhancing optimization techniques and LLMs to yield forthcoming generations of GenAI models characterized by heightened potency, user-friendliness, rapid trainability, and adaptability to specific projects or downstream applications. Adaptive GenAI-based tools, such as PrivateGPT \footnote{https://www.privategpt.io/}, an enclosed ChatGPT that encapsulates comprehensive organizational knowledge, have surfaced and garnered substantial community support. Another tool, MetaGPT \footnote{https://github.com/geekan/MetaGPT} illustrates a vision where a software development project can be fully automated via the adoption of role-based GPT agents.

For the first time in history, programming is becoming accessible to a broad spectrum of individuals, including those lacking a formal background in Software Engineering (SE). GenAI will offer opportunities to democratize coding, enabling individuals from diverse backgrounds to partake in software development. With its intuitive interface and robust automation capabilities, GenAI serves as a conduit for individuals to unlock their creative potential and manifest innovative concepts through coding, irrespective of their prior technical acumen. This inclusive approach to programming not only fosters a more diverse and equitable technology community but also initiates novel avenues for innovation and complex problem-solving within the digital era. 

The current AI landscape parallels the challenges inherent in edge machine learning, wherein persistent constraints related to memory, computation, and energy consumption are encountered. These hardware limitations are being effectively addressed through established yet efficacious technologies such as quantization and pruning. Concomitant with these hardware optimizations, there is a notable rise in community involvement, affording opportunities for research groups beyond the purview of major technology corporations. This augurs the imminent development of groundbreaking GenAI products finely tailored for the software industry.

\subsection{GenAI in Quality Assurance}%







 
\subsubsection{Historical Context}
Software Quality Assurance (SQA) activities are imperative to ensure the quality of software and its associated artifacts throughout the software development life-cycle. It encompasses the systematic execution of all the necessary actions to build the confidence that the software conforms to the implied requirements, standards, processes, or procedures and is fit for use\cite{planning2002economic}. The implied quality requirements include, but are not limited to, functionality, maintainability, reliability, efficiency, usability, and portability. Often, SQA and software testing are used interchangeably when software quality is concerned. However, both these terms are intricately intertwined\cite{planning2002economic}. While SQA is a broader domain, testing remains a key and integral part of it, executing the software to detect defects and issues associated with quality requirements. Testing is widely practiced in the software industry despite the fact that it is considered an expensive process.

\par The testing process begins with understanding the requirements and generating test case requirements thereafter. Depending on the software system under test, software development life-cycle in practice and implied requirements, various types of tests are usually conducted. These include but are not confined to (functional) unit testing, integration testing, system testing, acceptance testing, and finally, testing the (required) quality attributes as per the non-functional requirements.  

\par Through a recent and extensive survey\cite{testrail2023report}, several key challenges related to SQA and testing are disseminated by the SQA community. \textit{Development of automated tests, limited time, budget, and human resources, inadequacies of SQA training, adapting SQA to contemporary and evolving software development workflows, and understanding requirements before coding} are among the key challenges. A similar list of challenges is presented in another study\cite{liu4584792autotestgpt}. It reports that traditional methods of testing, and test case design consume a significant amount of energy, time and human resources. For Yuan et al.\cite{yuan2023no}, generating tests manually, according to the existing practices, is laborious and time-consuming. As a result, the efficiency of conducting these activities becomes low while the extent and intensity of these activities also rely on the experience of the SQA professionals\cite{liu4584792autotestgpt}. 

\par These challenges make it hard to meet the increasing demand for SQA and testing across software companies. On the other hand, recent advancements in GenAI, especially in automatic code generation through LLMs, demand a significant transformation in how SQA functions within the current paradigm. Reduced time and resources required to perform SQA functions by automating various activities are among the evident benefits of utilizing GenAI for SQA. In fact, one of the existing obstacles faced by the SQA community is the flaws in the test data that make automated test sets fail for software systems\cite{testrail2023report}. Utilizing GenAI in SQA to generate diverse test data based on specified requirements, scenarios, and use cases effectively addresses these challenges. Moreover, GenAI has the potential to generate insights from the analysis of existing test data, enabling the SQA community to produce more reliable software systems effectively and efficiently.  

\subsubsection{Key RQs}
In light of the challenges faced by the SQA community, we propose several key questions for using GenAI in software quality. The questions include:
\begin{enumerate}
    \item To what extent GenAI can replace developers in code reviews?
    \item Can GenAI help to produce test cases and test scenarios from Software Requirements Specification (SRS)?
    \item Are there common error patterns of AI-generated code?
    \item How GenAI can be integrated effectively into quality assurance pipelines?
    \item How can GenAI be utilized to automate acceptance criteria from high-level requirements?
    \item How can GenAI be used to aid the creation of diverse and effective test suites for software testing?
    \item How can GenAI assist in maintaining and updating test cases in case of requirement changes?
    \item What strategies can be employed to ensure that the generated test cases by GenAI are reliable, effective, and representative of real-world usage scenarios?
\end{enumerate}
 \begin{figure}[!htb]
        \centering
        \includegraphics[scale=0.33]{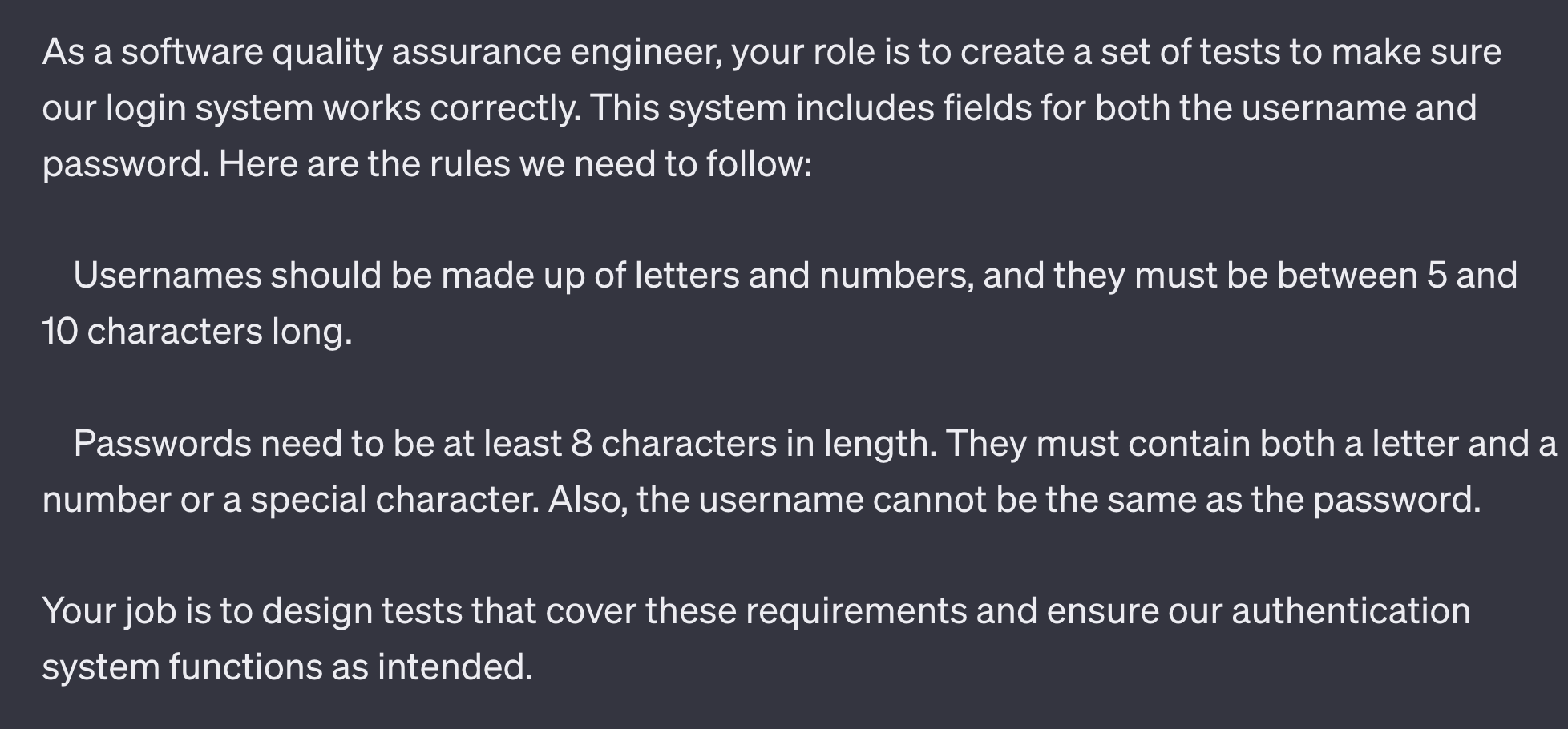}
        \caption[A Sample Prompt to Generate Test Cases Based on Requirements]{A Sample Prompt to Generate Test Cases Based on Requirements}
        \label{fig:t_c_prompt}
\end{figure}

\begin{figure}[!htb]
        \centering
        \includegraphics[scale=0.33]{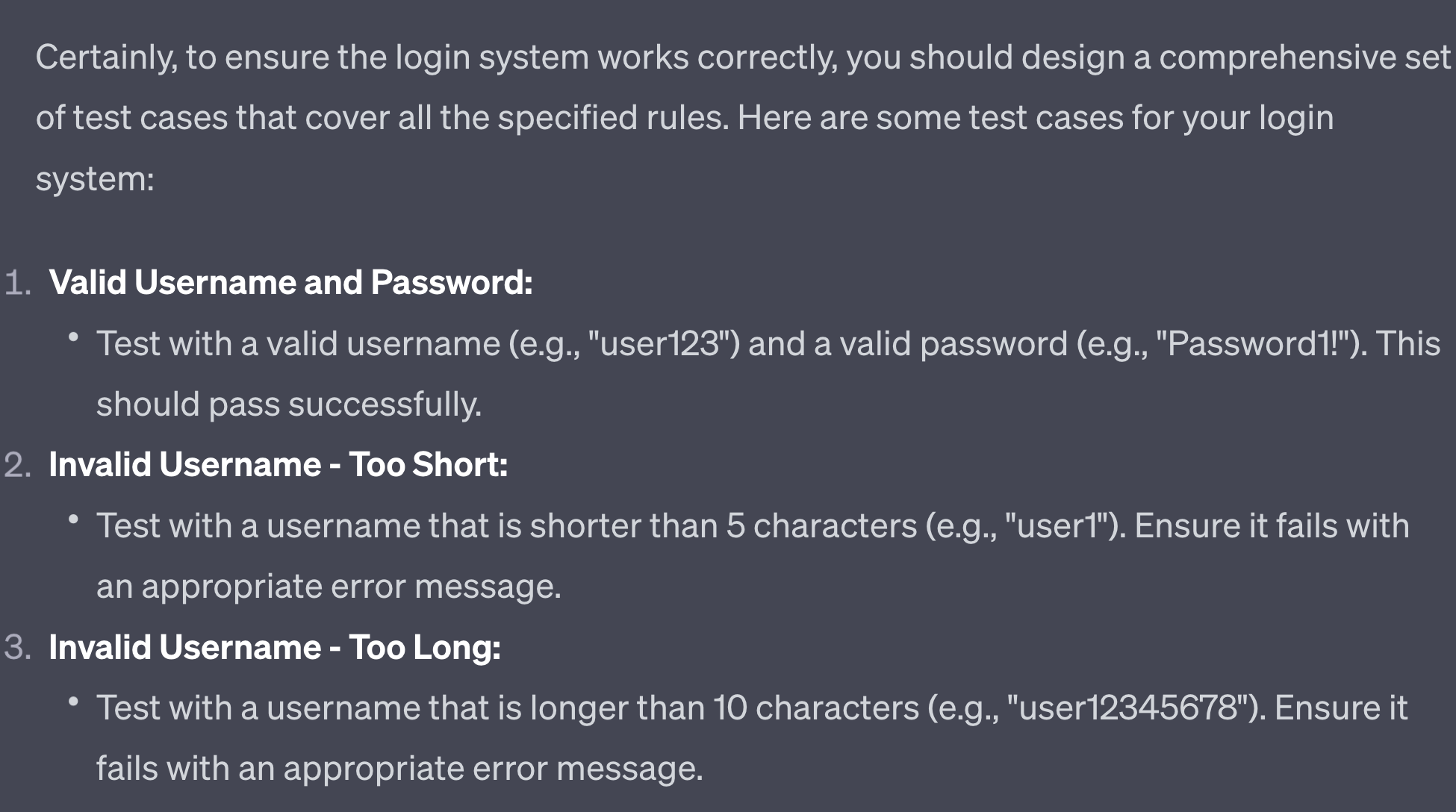}
        \caption[Part of the Generated Test Cases (Response)]{Part of the Generated Test Cases (Response)}
        \label{fig:t_c_response}
\end{figure}

\subsubsection{State-of-the-art}
Considering the challenges faced within this domain
, a large number of studies have been conducted to explore, propose and evaluate different QA approaches. The adoption of AI also has a long history in software quality research. Primarily, AI has been utilized in automating and optimizing tests, predicting faults, and maintaining software quality\cite{kotti2023machine}. With the recent advancement in AI models, GenAI might support various SQA activities. However, existing GenAI models (e.g. ChatGPT) are not particularly trained to perform these specialized jobs. Thus, they might offer limited performance when applied to particular situations\cite{jalil2023ChatGPT}. Nevertheless, these recent breakthroughs in the field have invited several researchers to find possible ways to address the challenges faced by the SQA community and tailor and evaluate existing models for SQA tasks. Regarding this, we report a number of studies that are aimed at understanding and evaluating the use and effectiveness of GenAI models in specialized settings.

\par Jalil et al. \cite{jalil2023ChatGPT} explored the use of GenAI to educate novice software test engineers. The authors reported that LLMs, like, for instance, ChatGPT have been able to respond to 77.5\% of the testing questions while 55.6\% of the provided answers were correct or partially correct. The explanations to these answers were correct by the of 53\%. With these percentages in mind, it has been ascertained that prompting techniques and additional information in queries might enhance the likelihood of accurate answers\cite{jalil2023ChatGPT}. Ma et al.\cite{ma2023scope} agreed with this by pointing out that the prompt design significantly impacts the performance of GenAI models. In this context, White et al. \cite{white2023chatgpt} proposed various prompt patterns to improve code quality and perform testing while interacting with GenAI models.  

\par Akbar et al. \cite{akbar2023ethical} reported the possible uses of GenAI models for software quality and testing. 
Yuan et al.\cite{yuan2023no} showed the potential of LLMs in unit test generation, which is among the most popular test phases in testing after model-based testing \cite{garousi2016systematic}. Similarly, it is also ascertained that GenAI models can be optimized to simulate various testing activities like, for instance, testing usability-related requirements of software systems\cite{akbar2023ethical}. As a result, software companies can save time and resources\cite{jalil2023ChatGPT}\cite{yuan2023no} and simultaneously improve the software quality and performance\cite{akbar2023ethical}. In addition, based on the literature survey, the study\cite{akbar2023ethical} also identifies several factors that can make an organization adopt GenAI models for testing and software quality. These factors include the generation of software test data, the capability to fine-tune these models for SQA tasks, generating test scenarios, bug reporting, automated reporting on the performance and software quality, and the tailored use of these models as testing assistants. Keeping these enablers in mind, we observed that the current literature provides no clear guidelines on how to achieve the objectives of using GenAI in SQA.

\par Liu et al.\cite{liu4584792autotestgpt}, in their recent study, attempted to leverage the capabilities of ChatGPT for testing purposes and developed a system for automated test case generation using ChatGPT. They report interesting comparisons between manually constructing test suites and generating them through ChatGPT. According to the experimental results of the study \cite{liu4584792autotestgpt}, the efficiency in generating test suits through their ChatGPT-based system exceeds 70\%. The new system uses ChatGPT as an intelligent engine to do the job. In contrast to  Liu et al.\cite{liu4584792autotestgpt}, Yuan et al. \cite{yuan2023no} claimed that only 24.8\% of tests generated by ChatGPT were executed successfully. The remaining tests suffered from correctness issues\cite{yuan2023no}\cite{wang2023software}. Most of these issues include compilation errors (57.9\%) followed by incorrect assertions (17.3\%). While on the other hand, Yuan et al. \cite{yuan2023no} found that the tests generated by ChatGPT are similar to human-generated tests with a sufficient level of readability and cover the highest test coverage. 

\par While discussing the scope of ChatGPT for code analysis, Ma et al. \cite{ma2023scope} used a significant number of codes and concluded that ChatGPT is great at understanding code syntax. However, it fails to fully comprehend dynamic semantics. The experimental evaluation of this study also demonstrates the GenAI model's limited ability to approximate dynamic behaviors.  

\par Lastly, in a large-scale literature review, Wang et al. \cite{wang2023software} collected and analyzed 52 empirical and experimental studies on the use of GenAI LLMs for software testing. The authors reported commonly employed LLMs, and software testing tasks that can be accomplished through LLMs. In this regard, test case preparation and program repair remained the two most commonly explored areas of prior investigations\cite{wang2023software}. Other areas include test oracle generation, bug analysis, debugging, and program repair. On the contrary, CodeX, ChatGPT, codeT5, CodeGen and GPT-3 are commonly utilized models by researchers to verify and validate particular but a few aspects of the software quality.  

\subsubsection{Challenges and Limitations}
Like software implementation, the utility of GenAI models for SQA activities is fully potential, but also challenging.  
The above studies show the illustrative use of GenAI in assessing code quality, analyzing both human and AI-generated code, conducting code reviews, and generating tests and test data. Nevertheless, inadequacies of current models do exist e.g. to assess the dynamic behavior of the code, human intervention to identify the hallucination, when dealing with code semantic structures and non-existent codes. Therefore, despite the promising benefits that GenAI offers to SQA, there is a need to evolve the existing methodologies, tools, and practices.
We present some of the challenges and limitations of GenAI for SQA extracted from the literature review: 
\begin{itemize}
    \item The tests generated by GenAI models currently demonstrate issues related to correctness, hallucination, and a limited understanding of code semantics\cite{liu4584792autotestgpt}\cite{ma2023scope}\cite{Zheng_2023}. These challenges are particularly notable in comprehending the dynamic behavior of the code. In addition, these challenges highlight the need for tremendous training data to refine models to perform specialized SQA tasks\cite{akbar2023ethical}\cite{zheng2023towards}. The training requires investing a significant amount of resources, which might not be a feasible option for all software development companies\cite{feldt2023towards}. These issues make the limited practicality of GenAI models in their current state, within the SQA context. 
    \item  Biasness in the training data may also affect the potential behavior of GenAI models towards SQA activities\cite{akbar2023ethical}. As a result, human intervention for manual correction or identification of vulnerabilities remains mandatory. Alongside this, biases in the training data might raise ethical issues or lead to a negative impact on software quality. 
    \item Output of GenAI models is highly associated with the prompting techniques\cite{jalil2023chatgpt}\cite{white2023chatgpt}. Providing more information or additional context of testing or quality might result in a better or improved answer. It further requires researchers to design proper prompt engineering techniques for SQA professionals.  
    \item Almost all of the existing studies on the topic(e.g. \cite{jalil2023chatgpt}\cite{liu4584792autotestgpt}\cite{ma2023scope}\cite{yuan2023no}) are mostly experimental studies and thus do not take into consideration the industrial context. Therefore, how GenAI models deal with real-world software quality issues remains a mystery. We need more real-world case studies and industrial examples to understand the effectiveness of various tasks of SQA with GenAI. Moreover, it would be interesting to study how well GenAI enhances the productivity of SQA professionals. 
\end{itemize}
\subsubsection{Future Prospects}
The existing studies stimulate the potential of utilizing GenAI in assessing software quality and thus demand reshaping the traditional SQA implementation methods, practices, and tools. Shortly, we foresee a significant number of studies within this domain. Therefore, based on the existing understanding of the use of GenAI in this domain and considering the challenges and limitations faced, we propose several possible avenues for forthcoming research studies:

   
\par \noindent\textbf{Exploring GenAI for Further SQA Tasks:} Future research might investigate conducting and evaluating early activities of SQA through GenAI. This includes creating a complete test plan or generating test requirements based on SRS. However, automating this process would still require a human expert to validate the outcome because these activities rely heavily on domain knowledge. The alternative way to achieve this could be training the GenAI models with relevant domain data\cite{wang2023software}. 
\par \noindent Likewise, much of the existing literature studied functional testing followed by security testing in terms of finding vulnerabilities and asking for bug repair\cite{wang2023software}. We found only one study \cite{akbar2023ethical} that alludes to the use of GenAI to test usability requirements. However, even in this case, the practices are not obviously described. Similarly, other areas of testing are not currently the focus of existing studies, i.e. acceptance testing, integration testing, and testing other software quality requirements like performance, reliability, efficiency, and portability. Therefore, future research on these SQA activities would be useful to show the effectiveness and limitations of existing GenAI models.   
\par \noindent\textbf{Exploring the Impact of Prompt Strategies :}
    Further research is also required to evaluate the effect of prompt strategies on GenAI to accomplish particular SQA tasks. The existing studies only employed four prompt engineering techniques so far, out of 12 \cite{wang2023software}. The effect of prompting on the GenAI models is also acknowledged by other studies e.g. \cite{white2023chatgpt}\cite{ma2023scope}\cite{akbar2023ethical}. Therefore, a future prospect might be to identify effective prompting strategies for a particular task, such as unit test code generation. 
    \par \noindent\textbf{Utilizing Latest GenAI Models:} Wang et al. \cite{wang2023software} found that Codex and ChatGPT are two largely explored GenAI models to assess the usefulness of GenAI for SQA activities. However, as expected, we do not find any study utilizing the latest version of ChatGPT i.e. ChatGPT 4.0. Therefore, investigating the use of the latest GenAI model might produce different performance results for varying SQA activities. Using a different dataset and a different GenAI model would offer a different effect on the performance results.

\subsection{GenAI in Maintenance and Evolution}

\subsubsection{Historical Context}
Software maintenance encompasses all the activities involved in managing and enhancing a software system after it has been deployed, ensuring the system's continued quality in responding to changing requirements and environments \cite{Przybylek_2018_EMSE}. As software maintenance is an expensive activity that consumes a considerable portion of the cost of the total project \cite{Jarzabek_2007,Przybylek_2018_EMSE}, software systems should be developed with maintainability in mind. Unfortunately, software maintainability is frequently neglected, resulting in products lacking adequate testing,  documentation, and often being difficult to comprehend. Therefore, to maintain a software system, one must usually rely on the knowledge embodied and embedded in the source code \cite{Przybylek_2018_EMSE}. In practice, software engineers spend a lot of time understanding code to be maintained, converting or translating code from one environment to another, and refactoring and repairing code. This is also an area with intensive research on the application of GenAI.

\subsubsection{Key RQs}
Excluding open challenges in code generation or quality assurance, which have already discussed in previous sections, we present below RQs that are specific to software maintenance tasks, i.e., code translation, refactoring, and code repair:

\begin{enumerate}
    \item How to improve code correctness in code translation tasks?
    \item  How can GenAI be leveraged to migrate an entire codebase to another language or framework?
    \item How can GenAI be leveraged to refactor a monolithic system into its microservices counterpart?
    \item How can GenAI support automatic program repair?
    \item How to integrate multiple LLMs or combine LLMs with specialized traditional ML models to leverage the collective strengths for automatic program repair?
    \item How can GenAI be leveraged to automate code smells and anomaly detection, facilitating the early identification and resolution of issues in software maintenance?
    \item How can GenAI be leveraged to automate bug triage?
    \item How can GenAI be leveraged to automate clone detection?
\end{enumerate}

\subsubsection{State-of-the-art}
The past year has witnessed significant transformations in software maintenance practice, largely catalyzed by the introduction of GitHub Copilot powered by OpenAI Codex. Leveraging LLMs can assist software maintainers in a wide range of activities, including code summarization, code translation, code refactoring, code repair, and more.

\emph{Code summarization} is the task of writing natural language descriptions of source code. Taking into account that according to several studies \cite{Anquetil_2007,Jarzabek_2007,Przybylek_2018_EMSE}, from 40\% to 60\% of the software maintenance effort is spent trying to understand the system, code summarization stands out as a pivotal technique for enhancing program comprehension and facilitating automatic documentation generation. It effectively addresses the frequent disconnect between the evolution of code and its accompanying documentation. Noever and Williams \cite{noever_chatbots_2023} demonstrated the value of AI-driven code assistants in simplifying the explanation or overall functionality of code that has shaped modern technology but lacked explanatory commentary. On a related note, Ahmed and Devanbu \cite{Ahmed_2022} investigated the effectiveness of few-shot training with ten samples for the code summarization task and found that the Codex model can significantly outperform fine-tuned models. Apart from writing new comments, automated code summarization could potentially help update misaligned and outdated comments, which are very common in SE projects\cite{Ahmed_2022}.

\emph{Code translation} involves the intricate process of converting code between different programming languages while preserving its functionality and logic. This task holds substantial practical value, especially in scenarios like migrating legacy code to modern technologies. For instance, billions of lines written in Cobol and other obsolete languages are still in active use, yet it is increasingly challenging to find and recruit COBOL programmers to maintain and update these systems. To bridge the gap between the available knowledge of individuals and the expertise needed to effectively address legacy problems, IBM recently introduced Watsonx Code Assistant for IBM Z mainframes. This tool helps enable faster translation of COBOL to Java and is powered by a code-generating model called CodeNet, which can understand not only COBOL and Java but also approximately 115 different programming languages.

\emph{Code refactoring} is the process of restructuring and improving the internal structure of existing source code without changing its external behavior. Madaan et al. \cite{Madaan_2023} explored the application of LLMs in suggesting performance-enhancing code edits. They observed that both Codex and CodeGen can generate such edits, resulting in speedups of more than 2.5$\times$ for over 25\% of the investigated programs in both C++ and Python. Notably, these improvements persisted even after compiling the C++ programs using the O3 optimization level. On a related note, Poldrack et al. showed that GPT-4 refactoring of existing code can significantly improve code quality according to complexity metrics \cite{poldrack_ai-assisted_2023}.

\emph{Code repair} refers to the identification and fixing of code errors that may cause program crashes, performance bottlenecks, or security vulnerabilities. Accelerating this process using LLMs is crucial, as developers invest a significant amount of engineering time and effort in identifying and resolving bugs that could otherwise be devoted to the creative process of software development. Automated program repair (APR) has been a topic of much interest for over a decade in the SE research community. Zhang et al. proposed a pre-trained model-based technique to automate the program repair \cite{zhang_boosting_2023}. Xia et al. \cite{Xia_2022} designed three different repair settings to evaluate the different ways in which state-of-the-art LLMs can be used to generate patches: 1) generate the entire patch function, 2) fill in a chunk of code given the prefix and suffix 3) output a single line fix. The findings suggest that directly applying state-of-the-art LLMs can substantially outperform all existing APR techniques. In contrast, Pearce et al. \cite{Pearce_2023} examined zero-shot vulnerability repair using LLMs (including Codex) and found that off-the-shelf models struggle to generate plausible security fixes in real-world scenarios. In a different context, Asare et al. \cite{Asare_2023} discovered that Copilot has the potential to aid developers in identifying potential security vulnerabilities. However, it falls short when it comes to the task of actually fixing these vulnerabilities. Lastly, Mastropaolo et al. \cite{Mastropaolo_2021} fine-tuned the T5 model by reusing datasets used in four previous works that used deep learning techniques to (i) fix bugs, (ii) inject code mutants, (iii) generate assert statements, and (iv) generate code comments. They compared the performance of this model with the results reported in the four original papers proposing deep learning-based solutions for those four tasks. The findings suggest that the T5 model achieved performance improvements over the four baselines.


\subsubsection{Challenges and Limitations}
As detailed in the preceding subsections, GenAI models demonstrate proficiency in maintenance tasks that require an understanding of syntax, such as code summarization and code repair. However, they encounter challenges when it comes to grasping code semantics \cite{Zheng_2023,Hou_2023}. Consequently, the performance of GenAI is still limited when it comes to several realistic scenarios \cite{Li_2023}.

Moreover, applying available GenAI tools to code translation tasks also faces challenges and limitations. Primarily, the amount of source code that LLMs can consider when generating a response is constrained by the size of the context window. To illustrate, ChatGPT Plus imposes a token limitation of 8,000 characters, equivalent to roughly 100 lines of code. Consequently, when dealing with extensive codebases, a straightforward approach of inputting the entire source code at once becomes unfeasible. For instance, in the case of GPT Migrate \footnote{https://github.com/0xpayne/gpt-migrate}, code migration is a solvable problem if the entire codebase is small enough to fit in the context window. However, addressing larger codebases remains an unresolved issue. As this restriction on code length is inherent in the transformer architecture, overcoming this limitation would necessitate a significant breakthrough in the field. Besides, acquiring a large amount of high-quality source code and target code pairs is challenging, which may limit the model’s learning and generalization capabilities. 
Finally, as even minor translation errors can render the generated code non-functional, code correctness and precision are of paramount importance, posing challenges for LLMs.

\subsubsection{Future Prospects}
At the current stage, LLMs can serve as useful assistants to software maintainers, but they are very far from replacing completely humans \cite{Zheng_2023}. Nonetheless, LLMs are continuously advancing, and they still hold significant untapped potential. One promising avenue is the concept of Collaborative LLMs \cite{Hou_2023}. This approach involves the integration of multiple LLMs or their combination with specialized ML models to minimize the necessity for human intervention. In this setup, multiple LLMs function as distinct "experts", each handling a specific subtask within a complex task \cite{Dong_2023}. By harnessing the collective strengths of diverse models, more precise and efficient outcomes can be achieved \cite{Hou_2023}. Furthermore, regarding the enhancement of correctness and performance in LLMs for vulnerability detection, future research should focus on combining LLMs with formal verification techniques, incorporating previously related edits, or implementing self-debugging algorithms \cite{Zheng_2023}.

\subsection{GenAI in Software Processes and Tools}

\subsubsection{Historical Context.} Software processes are essential components of software development, helping developers create, maintain, and deploy software applications efficiently and effectively. Industry continues to look for novel ways of improving SE processes. So far, common software processes include Waterfall, Agile, Continuous Sofware Engineering and Lean Software Development \cite{sommerville_software_2015,nguyen-duc_towards_2016,fitzgerald2017continuous}. 
The implementation of software processes often goes hand in hand with the adoption of various tools. For instance, Agile projects utilize Kanban-based management tools, and Continous Software Engineering with Continuous to automate the build, testing, and deployment of software, enabling rapid and reliable releases. As the latest movement, ML/AI for SE and specifically GenAI is receiving widespread interest in this regard. While early pioneers have been investigating and employing ML in SE even before the popularity of systems like ChatGPT, the integration of GenAI into SE practices has emerged as a critical industry priority. Although individual developers have started embracing these technologies for personal assistance\cite{feng_investigating_2023}, the endorsement and implementation of GenAI tools at the company level by management in many organizations are lagging.

\subsubsection{Key RQs.} In this regard, the main question is \textit{what} to do with GenAI, and \textit{how} to do so effectively while considering the current working processes, practices and environments. For example, a practitioner looking into Copilot could ask which tasks are the tool suited for and what are the good and best practices for utilizing them for said tasks. Given the versatility of GenAI, the potential applications of these tools in SE are still being explored. To this end, as GenAI tools are used to support and automate tasks by different personnel, the question of how they impact productivity becomes increasingly relevant from a business point of view. In the long run, the question of how these tools will transform SE processes and what types of new processes are born due to their use is important. We highlight specific RQs related to the SE process as below:

\begin{enumerate}
    \item  What strategies can be developed to enable GenAI models to support continuous integration and deployment (CI/CD) practices?
    \item How can different GenAI tools be incorporated into software development environments?
    \item To what degree do GenAI tools improve developer productivity?
    \item How can GenAI be utilized to aid in the automation of code commit, pull, and push requests?
    \item What approaches can be used to enhance the integration of GenAI into development environments, providing real-time intelligent assistance and feedback during code editing and development activities?
    \item How can GenAI support the automation of build and deployment processes?
    \item What is the impact of GenAI tools on developer productivity in various software development tasks and across different levels of developer seniority?
    \item How can GenAI contribute to Citizen development?
    \item How will GenAI tools transform fast-faced software development processes, i.e. Agile, Continuous Development and Lean Development?
\end{enumerate}

\subsubsection{State-of-the-art.} There currently exist a few studies focusing on GenAI and software processes. The tertiary study by Kotti et al. \cite{kotti_machine_2023} provides an overview of AI use in SE overall, reconsidering deficient SE methods and processes. Weisz et al. \cite{weisz_perfection_2021} and Whites \cite{white_chatgpt_2023} explore the applications of GenAI in programming-related tasks, while in another work, Whites and co-authors also discuss prompt patterns for SE-related prompts \cite{white_prompt_2023}. Ross et al. proposed a programmer assistant based on LLMs, and explored the utility of conversational interactions grounded in code \cite{ross_programmers_2023}. The impacts of the utilization of GenAI tools have also been discussed, primarily from the point of view of productivity \cite{ziegler_productivity_2022}. While some existing papers speculate on how SE might be transformed by (Gen)AI going forward, this remains on the level of speculation, although Bird et al. \cite{bird_taking_2023} provide some insights into how the use of GitHub Copilot changes the way software developers work. 

Software processes are more often presented as a part of study contexts. Petrovic et al. presented a use case where ChatGPT can be used for run-time {DevSecOps} scenarios \cite{petrovic_machine_2023}. In a game development context, Lanzi et al. proposed a collaborative design framework that combines interactive evolution and large language models to simulate the typical human design process \cite{lanzi_chatgpt_2023}. Ye et al. investigated the impact of ChatGPT on trust in a human-robot collaboration process \cite{ye_improved_2023}.

\subsubsection{Challenges and Limitations.} While many companies are interested in the potential prospect of automation and the resulting productivity and financial gains, many existing papers highlight the need for human oversight in utilizing GenAI in SE processes due to the potential errors GenAI tools make (e.g., programming \cite{nascimento_comparing_2023}). In this regard, Moradi et al. \cite{moradi_dakhel_github_2023} argue that tools such as GitHub Copilot can be assets for senior developers as supporting tools but may prove challenging to more novice developers who may struggle to evaluate the correctness of the code produced by the AI. Becoming overconfident in GenAI tools may result in a lack of criticism towards the code produced \cite{imai_is_2022}. Overall, good practices in utilizing GenAI in the SE process are still missing, and only a number of early studies have begun to inspect the practical applications of GenAI in the SE process. Though the promise of code generation with GenAI is now largely acknowledged and, to a large extent, demonstrated in practice as far as market viability is concerned by the explosive growth of the GitHub Copilot service, the utilization of GenAI for other SE processes is far less explored, and even the impact of GenAI on code generation remains poorly understood (e.g., productivity in different contexts).

\subsubsection{Future Prospects.} GenAI tools, alongside AI in general, have the potential to transform SE processes as we know them today. It is argued that, in the near future, the work of developers will shift from writing code to reviewing code written by AI assistants \cite{bird_taking_2023}. Carleton et al. \cite{carleton_architecting_2022} argue that the key is "shifting the attention of humans to the conceptual tasks that computers are not good at (such as deciding what we want to use an aircraft for) and eliminating human error from tasks where computers can help (such as understanding the interactions among the aircraft’s many systems)". Such larger, fundamental shifts in organizing SE processes are arguably at the center of the future prospects for the SE process research area. For example, while the changing role of developers is acknowledged, with AI tools making it possible for developers to automate menial tasks in order to focus on tasks that require more expertise, what exactly these tasks might be in the future remains an open question.


\subsection{GenAI in Engineering Management}

\subsubsection{Historical Context}- Engineering management is the SE branch that deals with managing principles and practices concerning the engineering field. The critical aspects of this field include project management \cite{eisner}, human resources \cite{tambe}, team, leadership \cite{amm}, technical expertise, budgeting, resource allocation, risk management \cite{kham}, communication, quality assurance, continuous improvement, ethics, and professionalism \cite{blanchard2004system}. A project undertaken under engineering management ensures that it is completed successfully, on time, within budget, and with high-quality standards \cite{smith2002engineering}. Crucial is to manage the project under several laid out regulations \cite{blanchard2004system}. Tambe et al., in Human resource management and AI, identified four types of challenges. First, complex phenomena surround the human resource. Second, the chain that bounds the small data sets, i.e., the people working in an organization, is minor in comparison to the purchases made by the customer. Third, ethical obligations are involved with legal constraints. Fourth, the response or behaviors of employees towards the management \cite{tambe}. Wang et al. proposed a systematic roadmap to navigate the next step of AI in product life cycle management. The study further mentions the advantages and challenges of applying AI in product cycle management. \cite{wang2021artificial}.\\

\subsubsection{Key RQs}- We consider the following RQs that present some open challenges in managing software development projects:
\begin{enumerate}
    \item What are the implications for team sizes and dynamics of the use of GenAI in software development?
    \item Can a trained AI tool speed up the enrolment process of a new employee?
    \item  To what extent should experience developers be retrained to adapt to GenAI use in their work?
    \item How can organizations foster a culture that promotes the symbiotic collaboration between software developers and GenAI systems, creating an environment that encourages trust, learning, and innovation?
    \item What could be an efficient human-ai collaboration workflow?
    \item Do different SE roles feel a stigma about using AI assistants to help them with their jobs?
    \item What is the economic impact of incorporating GenAI in software development processes, considering factors such as cost savings, productivity gains, and overall return on investment?
\end{enumerate}

\subsubsection{State-of-the-art}
For project management in general, Holzmann et al. conducted a Delphi study on 52 project managers to portray the future AI applications in project management \cite{holzmann2022expectations}. The author presented the list of most and least important project management areas where GenAI can be adopted. Prifti et al. identified the pain points or problems in project management. Then, the paper proposes how AI can play a role and help optimize weak points. AI's assistance helps project managers become more productive and skilled in organizations \cite{prifti2022optimizing}. 

In software project management, AI-based approaches can be useful for various project estimation and allocation tasks. Dam et al. explored the effort estimation problems in an Agile context \cite{dam2019towards}. Elmousalami analyzed and compared the 20 AI techniques for a project cost prediction \cite{elmousalami2020comparison}. Fu et al. proposed a GPT-based Agile Story Point Estimation approach, highlighting a practical need for explanation-based estimation models \cite{fu_gpt2sp_2023}. Alhamed evaluated the Random Forest and BERT model in software ask effort estimation. \cite{alhamed_evaluation_2022}. The authors suggested that BERT shows marginally improved performance than other approaches.  
\begin{table}
    \centering
    \begin{tabular}{|p{7cm}|}
    \hline
        Most important PM functions for GenAI adoption \\\hline
            \begin{itemize}
             \item Create project schedule
             \item Analyze implications of missing deadlines
             \item Create WBS and tasks list
             \item Create project budget
             \item Identify scope creep and deviation
             \item Produce dynamic risk map
             \item Extract deliverables
             \item Update project progress and schedule
             \item Prioritize tasks
             \item Allocate team
         \end{itemize}
        \\
        \hline
    \end{tabular}
    \caption{A list of relevant and important PM tasks can be supported by AI tools \cite{holzmann2022expectations}}
    \label{tab:my_label}
\end{table}

To contribute to the discussion about GenAI in an industrial context, Chu explored the impact of ChatGPT on organizational performance (productivity), reporting various use contexts but no notable productivity increases \cite{chu_assessing_2023}, although without a specific focus on SE-related work. Ebert \& Louridas \cite{ebert_generative_2023} reported industrial, empirical results from a case study, discussing the overall promise of GenAI for software practitioners and the current state of the field, and highlighting potential application contexts in SE \textit{(media content creation and improvement, generative design, code generation, test case generation from requirements, re-establishing traceability, explaining code, refactoring of legacy code, software maintenance with augmented guidance, and improving existing code, among others)}.

\subsubsection{Challenges and Limitations}
In this SE area, the lack of industrial studies, and especially the lack of studies with an organizational focus, presents as one of the notable gaps. Pan et al. presented a critical review of AI-generated future trends in construction engineering and management. The study outlines six crucial concerns regarding AI in construction engineering, which are (1) knowledge representation, (2), information fusion, (3) computer vision, (4) natural language processing, (5) process mining, and (6) intelligence optimization \cite{pan2021roles}. Another study described the future challenges of adopting AI in the construction industry \cite{abioye2021artificial}, presenting several challenges. They are the cultural issues and explainable AI, the security concerns, the talent shortage of AI engineers, the high cost in the construction industry, ethics, governance, and finally, the computing power and internet connectivity that lead to construction activities. There are several challenges concerning software project management, too, such as needing more understanding concerning the software development team and awareness of the project. The project context could be complex, with limited knowledge available that might create developers in the group with biased decisions. In other words, biased data created could lead to subjective decisions that result in less effective software projects \cite{parikh2023empowering}.  

\subsubsection{Future Prospects}
AI is regarded as an originator, coordinator, and promoter of innovation management. At the company level, AI applications can be used for opportunities like searching for information, generating ideas, and value creation. In terms of organization design, AI could be beneficial in looking into the structure of the firm, customer engagement, and human decision-making processes. AI can be useful for generating information to data access that could help at the startup and SME levels \cite{brem2021ai}. In software project management, AI can be very handful in assisting the organization of resources and allotting, managing the risks involved in the projects, automation of the task allocation, estimating the cost of the project and its related requirements and planning the agile sprints \cite{dam2019towards}. Monitoring the performance, capturing the problems inside the project, providing support suggestions, and documenting the project could make some prospects replaceable by AI \cite{song2023artificial}.


\subsection{GenAI in Professional Competencies}



\subsubsection{Historical context.}
There are skill shortages and talent gaps~\cite{humanskillshortage,turkeyskillgap,skillgaptraining} in the software industry: this situation has been going on for a while and companies struggle to find qualified employees~\cite{womensoftware,jiagui_swot_2023}. On top of that, there is a gender gap in technology-related study fields~\cite{womensoftware,genderinstem} and a soft skills gap~\cite{softskillsgap,humanskillshortage} that make it even harder to find professionals and have balanced teams able to detect and correct unfair biases, e.g., caused by the selection of data sources~\cite{datagenderbias}. The question then is how to offload repetitive tasks from existing developers by smoothly integrating GenAI into their workspace. On the one hand, as happened with previous technologies empowering developers, integrating GenAI into the workspace and ensuring that developers smoothly work alongside GenAI tools will probably require training~\cite{skillgaptraining}. However, as seen in previous research, some users might already have an intuition on how to work with them, maybe depending on their seniority level or background~\cite{turkeyskillgap,skillgaptraining}. On the other hand, GenAI could also be part of the training of software developers of the future, partially addressing the aforementioned skill shortages, as discussed in Section~\ref{sec:education}.

\subsubsection{Key RQs.}
Not only academic research teams are studying how to effectively integrate GenAI within Agile teams to increase their productivity, also companies are actively exploring how these technologies will affect the work dynamics of their employees and, more specifically, of developers. Future research should explore the integration of GenAI tools in professional practices, covering productivity impacts, necessary skills and training, addressing skill shortages, GenAI as an advisor, and perceptions of integration strategies in Agile software development. These high-level concerns are addressed by the five suggested RQs listed below:

\begin{enumerate}
    \item How can GenAI address the skill shortages and talent gaps in the software industry, and how would different strategies affect the desired profile for future software developers?
    \item How would various strategies for GenAI integration within Agile software development, such as in up-skilling and re-skilling professionals, offloading repetitive tasks, or human resource management, be perceived?
    \item What is the impact of GenAI tools on productivity across various software development tasks, with respect to developer seniority and at the organizational level at which the GenAI interventions take place (i.e., team or team member)?
    \item How can GenAI be effectively used as an advisor to enhance the productivity and skills of software professionals, and what are the key challenges and opportunities in designing training programs for this?
    \item What skills and training are necessary for software engineers to generate quality outputs using GenAI and effectively work alongside GenAI systems? Does the level of SE training correlate with these goals?
\end{enumerate}

These RQs reflect the main lines of research from industry and academia in understanding how GenAI tools could mitigate the current issue of finding software professionals and either offload or up-skill existing developers to strengthen their teams. 

\subsubsection{State-of-the-art}
As previously stated, both academic and industrial researchers are actively exploring how GenAI technologies will affect the work processes and dynamics. An increasingly adopted tool for such explorations is GitHub Copilot~\cite{github-inc}. Copilot generates pieces of code given prompts in the form of comments or function signatures thanks to its training with open-source projects hosted by the GitHub platform. Other platforms include local versions of GPT-models~\cite{openai2023gpt4} or BARD~\cite{bard}, both based in LLMs. For these tools, however, it is unclear whether their usage violates the licenses and internal best security practices. As a result, informed human oversight will still be needed to ensure the trustworthiness of the generations~\cite{AIact}.



\subsubsection{Challenges and limitations.}
Software development companies are often cautious about granting external access to their systems due to strict policies safeguarding proprietary information and sensitive data. Convincing companies to participate in research on GenAI interactions can be challenging, even though demonstrating the mutual benefits of testing the integration of GenAI tools in their unique working environment can increase their willingness to collaborate.
As in any other research endeavour involving human participants, obtaining informed consent and maintaining confidentiality and anonymity are crucial. However, in this case, research results may offer insights into the different productivity or work dynamics of teams and team members. Therefore, researchers must ensure that their findings are put into context and that the validity of their conclusions is clearly stated. This precautionary approach is essential to prevent companies from making drastic decisions based on potential misinterpretations of the research results.

\subsubsection{Future prospects.}
In conclusion, addressing these RQs and carefully applying the results to various work environments and domains can serve as a valuable guide for the sensible integration of GenAI tools within Agile software development. This integration should consider the progressive adaptation of work dynamics and their adoption by the different stakeholders. The answers to these questions will, on the one hand, prepare companies and industrial players for the new technological setup where software developers can be offloaded of repetitive tasks. On the other hand, this knowledge will play a crucial role in shaping the integration of GenAI into educational and re-skilling activities for future software developers, as discussed in Section~\ref{sec:education}, indirectly contributing to mitigate the present talent gap~\cite{humanskillshortage,skillgaptraining}.


\subsection{GenAI in Software Engineering Education} \label{sec:education}

\subsubsection{Historical Context.}
The development of engineering education has been inextricably linked to the technological development of the era. Technological adoption for better education is always an important topic in higher education. SE research has extensively looked at the use of chatbots to support students outside the classroom. These chatbots can provide automated assistance to students, helping them with coding and problem-solving tasks [7], providing explanations and examples, as well as helping students identify errors in their solutions \cite{gonzalez_improving_2022,Verleger2018,binkis_rule-based_2021,Ismail2019,Carreira2022}.

Another popular SE education paradigm is an automated assessment of exercises. Automated assessment techniques can help provide students with immediate feedback on their work, allowing them to quickly identify and correct errors \cite{paiva_automated_2022}. This approach can also facilitate self-assessment and self-examination among students, enabling them to explore various solutions and identify areas where they require further improvement \cite{cukusic_online_2014}. Another active research area that is gaining attention is how learning can be individualized. Current AI approaches focus on finding questions of similar difficulty to a given question \cite{qadir_engineering_2023}.

\subsubsection{Key RQs}
GenAI has the potential to revolutionize the existing education landscape by advancing the three above trends. For future research actions, there are several interesting themes that can be framed as RQs, as listed below:
\begin{enumerate}
    \item How can GenAI be effectively integrated into SE education curricula to enhance student learning outcomes and develop future-ready skills?
    \item  In a classroom environment with a “virtual” assistant with the capacity of i.e. ChatGPT, how can it be set up to effectively achieve the learning objectives?
    \item How can GenAI be used to personalize and adapt SE education to cater to diverse student needs, learning styles, and skill levels?
    \item How does the integration of GenAI in SE education foster critical thinking, problem-solving skills, and computational creativity among students?
    \item What assessment methods and tools can be developed to evaluate students' understanding, competency, and proficiency in GenAI concepts and applications?
    \item What are the long-term impacts of including GenAI in SE education, such as its influence on students' career paths, job prospects, and contributions to the technology industry?
    \item In what activities the use of AI can harm the learning outcome that the participants have?
    \item What are the ethical considerations and implications of teaching GenAI in SE education, and how can educators promote responsible AI usage and ethical decision-making among students?
    \item What are the challenges and opportunities in training IT educators to effectively incorporate GenAI concepts and tools into their teaching practices?
\end{enumerate}

\subsubsection{State-of-the-art}
We found a large number of studies discussing the potential of GenAI in education in general and in SE education, in particular, \cite{daun_how_2023,bull_generative_2023,berrezueta-guzman_recommendations_2023,banic_pair_2023,jalil_chatgpt_2023,ashraf_chatgpts_2023}. Bull et al. interviewed industry professionals to understand current practices, implying a visionary future of software development education \cite{bull_generative_2023}. Jalil et al. explored opportunities for
(and issues with) ChatGPT in software testing education \cite{jalil_chatgpt_2023}.

There also emerges empirical evaluation of GenAI in different teaching and learning activities. Savelka et al. analyzed the effectiveness of three models in answering multiple-choice questions from introductory and intermediate programming courses at the post-secondary level \cite{savelka_large_2023}. A large number of existing papers discuss the potential challenges GenAI presents for educators and institutions, which is also a recurring topic in these SE papers. However, these papers also discuss the potential benefits of GenAI for SE education. Yilmaz et al. conducted experiments with their students and found that ChatGPT could enhance student programming self-efficacy and motivation, however, it is important to provide students with
prompt writing skills \cite{yilmaz_effect_2023}. Philbin explored the implications of GenAI tools for computing education, with a particular emphasis on novice text-based programmers and their ability to learn conceptual knowledge in computer science. The author presented many positive angles on GenAI adoption and suggested the integration of basic AI literacy in curriculum content to better prepare young people for work \cite{philbin_exploring_2023}

\subsubsection{Challenges and Limitations} 

While GenAI tools like ChatGPT offer a revolutionary approach to facilitating SE education, they are not without limitations. A recent paper presented the perceived weaknesses of ChatGPT in nine SE courses. The authors presented the weaknesses in three categories: theory courses, programming courses, and project-based courses. 

In a theory course, a common challenge of ChatGPT is that its responses are limited to the information provided in the question. This means that if a question is poorly formulated or lacks relevant details, the response from ChatGPT may not be satisfactory. To use ChatGPT effectively, students need to be prepared with critical and innovative thinking, which does not yet exist in any college's curriculums. The problem of the non-deterministic nature of AI is particularly significant in an educational context, as different answers can be generated for the same question, which could lead to confusion for students. Besides, ChatGPT has limited domain knowledge, so questions that require a lot of context or background information may not receive meaningful answers without sufficient input from the user. 

Regarding programming courses, while GenAI tools can provide information and guidance on programming concepts, students cannot develop their practical skills solely by using ChatGPT. This applies to complex programming concepts, intricated problems, especially those related to integrating libraries or, specific system configurations, or large software systems. AI tools sometimes offer deprecated solutions or code snippets, meaning they are outdated or no longer recommended for use. Using deprecated solutions can lead to inefficiencies or conflicts in modern coding environments. Moreover, with lectures with a lot of visual representations, ChatGPT relies on natural language processing and may struggle to interpret visual information. For example, in a lecture about 3D modeling, ChatGPT may not be able to explain a particular aspect of the modeling process that is best understood through visual demonstration. While ChatGPT can provide information on general programming concepts, it may not be able to provide insight into specific practices used in a particular industry. For example, a student studying game development may need to learn about specific optimization techniques used in the gaming industry, which ChatGPT may not be familiar with. Furthermore, as technology and programming languages evolve rapidly, AI's knowledge might lag behind the most recent advancements or best practices. Some students pointed out that e.g. ChatGPT lacked information about the latest developments, affecting its ability to provide the best or most relevant answers. Furthermore, AI models don't possess an inherent understanding of context. There were situations where the tool couldn't recognize missing methods in a Repository class or provide solutions based on a broader project context, which a human tutor might be able to grasp.

Regarding project-based learning, where project contexts can change frequently, it is necessary to update the context information for the tool to catch up with project progress. For example, if a project in a software engineering class changes its requirements, students may need to make changes to their code. ChatGPT may not be able to keep up with these changes and may not be able to provide relevant advice to students. This is also because real-world settings can not be fully described as input for the tool. ChatGPT is based on pre-existing data and may not be able to provide practical advice for unique situations that occur in real-world settings.

\subsubsection{Future Prospects}

In conclusion,  GenAI tools, like ChatGPT, have shown significant promise in SE education. In the future, We foresee a comprehensive integration of GenAI tools at different levels for both educators and learners. Future research will address today's gap to visualize this scenario:
\begin{itemize}
    \item \textbf{The integration of GenAI into SE teaching content and approaches}. There is a pressing need for educational institutions to adapt their curriculum and teaching methodologies to incorporate GenAI concepts, tools, and practices. Concerning curriculum development, it becomes essential to determine which GenAI-related knowledge, such as prompt engineering, AIOps, and AI-human interaction, should be integrated into existing lectures on core SE topics like requirements engineering, software design, software testing, and quality assurance. On the teaching approach, how should such integration be done? GenAI tools like CoPilot, and ChatGPT also present a great potential to assist both educators and learners. There is a need to develop methodologies that help educators and AI developers work collaboratively to ensure that AI recommendations and adaptations align with established learning outcomes and instructional strategies. Besides, while AI can leverage large volumes of data to personalize learning experiences, there is a lack of standardized methodologies for collecting, processing and utilizing educational data ethically and effectively.
    \item \textbf{A new assessment approach for student-GenAI collaboration} A significant knowledge gap exists concerning the development and implementation of a novel assessment approach tailored to evaluate the efficacy and outcomes of student-GenAI collaboration in educational settings. We can not assess the student project delivery without considering the level of CoPilot involved in their project. Research is needed to create fair and reliable assessment techniques that differentiate between student-generated and AI-generated work. Project requirements and learning objectives might need to be updated as well to ensure that evaluations accurately reflect students' abilities and knowledge
    \item \textbf{An evaluation of the impact of AI-enabled curriculum and teaching approaches on learning outcome and student engagement} It is necessary to delve into empirical studies to measure and evaluate the impact of GenAI-related content and teaching approaches on student learning outcomes and engagement. Research in this area can involve conducting controlled experiments and collecting data to quantify the effectiveness of AI-assisted approaches compared to traditional methods. Assessment needs an ethical guideline focusing on how GenAI is used in SE courses.
\end{itemize}

\subsection{GenAI in Macro aspects}


\subsubsection{Historical context}
 The use of GenAI for SE brings implications and opportunities to other spheres beyond the technical aspects, such as market, business, organizations and society. This phenomenon can be observed with other technologies. For example, the advent of cloud computing, in which computing resources are hired on demand without the need for large upfront investments in infrastructure, allowed the cheap experimentation of new digital products~\cite{Ewens2018}, paving the way for methodologies for the development of startups focused on quick feedback from customers, such as Customer Development~\cite{Blank2007} and Lean Startup~\cite{Ries2011}. A more recent example is the emergence of more powerful AI algorithms and their reliance on data, leading to a growing research interest in the ethical aspects of this use~\cite{Vakkuri2020}, i.e., AI ethics.

\subsubsection{Key RQs}

We present the key RQs identified regarding macro aspects as listed below:
\begin{enumerate}
    \item How can GenAI be employed to support novel and innovative software-intensive business models?
    \item How can the adoption of GenAI in software development affect the competitiveness and market positioning of software-intensive companies, and what strategies can be employed to leverage AI for business growth?
    \item What are the potential barriers and challenges to the adoption of GenAI in software-intensive businesses?
    \item How does the integration of GenAI in software-intensive businesses affect software licensing models, and intellectual property rights?
    \item How can software businesses effectively navigate the ethical considerations and societal impacts of using GenAI?
    \item How does the adoption of GenAI impact the sustainability of software businesses?
\end{enumerate}

The first aspect included in this section regards the business of software. Any new technology brings several economic opportunities usually exploited by new companies. The application of GenAI to software development could bring several innovation opportunities for the players involved. A better understanding of this process, including a clear view of its associated challenges, would help software startups and established companies create and capture value for their customers by developing novel or adapting existing business models effectively. Still, regarding economic aspects, it remains unclear how the fact of AI creating intellectual property (IP) could change the management of this IP, including protection, use, and monetization. The second aspect regards the implications of the use of GenAI in software development to the environment and society. For example, GenAI adoption might reduce the need for human software developers, leading to a large contingent of job losses and the consequent need for retraining these people. Another issue might be the environmental impact of GenAI given, for example, the amount of resources needed to train the models.

\subsubsection{State-of-the-art}
Moradini conducted a narrative review to analyze research and practice on the impact of AI on human skills in organizations \cite{morandini_impact_2023}. The authors suggested the identification of needed skills for the adoption of AI and provided training and development opportunities to ensure their workers are prepared for the changing labor market. Kanbach et al.~\cite{Kanbach2023} performed a business model innovation (BMI) analysis on the implication of GenAI for businesses. The authors identified some propositions based on the impact on innovation activities, work environment, and information infrastructure. In particular, they discussed the SE context identifying possibilities for value creation innovation, e.g., automatic code generation, new proposition innovations, e.g., AI-powered software development platforms, and value capture innovations, such as improving operational efficiency by, for example, task automatization. On the other hand, there have been some calls for action regarding the regulation of GenAI, for example, in the healthcare sector~\cite{Mesko2023}.


\subsubsection{Challenges and Limitations}

As essentially interdisciplinary aspects, the research on these topics will have to converge to other fields, such as economics, psychology, philosophy, sociology, and anthropology. For example, concerns regarding intellectual property involve not only economic but also more fundamental aspects, such as what a person is, that are the study field of philosophers. These aspects are also associated with national and supranational politics and economics. Different countries might decide to create specific legislation or public policies depending on their political and economic tradition. 

\subsubsection{Future Prospects}

GenAI is an emerging technology that will develop even further in the following years. As it becomes increasingly capable of performing SE-related tasks, its implications on the macro aspects discussed above will grow. GenAI will be increasingly employed as part of software-intensive business models or even to suggest modifications to these business models. This process will represent a challenge for consolidated companies but also an opportunity for new incumbents. Current market leaders will need novel strategies to cope with these challenges and innovate, and novel software-intensive business models will emerge. In the following decades, we will probably see the downfall of current juggernauts and the emergence of new ones. Not only companies but also individuals will be affected, increasingly being replaced by AI. In this regard, we will probably watch a change in labour force, including professions related to software development, towards new careers, many of these initiated by GenAI, in a similar movement that developed countries experienced after the deindustrialization and the transfer of labour force to the services segment. Finally, we will probably see in the following years the emergence of new legislation regulating the use of GenAI and the products created by it. This phenomenon could lead to tensions among different actors in society, such as the strike of writers in Hollywood that paralyzed the production of several movies and TV series, motivated, among other things, the use of GenAI for script generation. Similarly, it could lead to international tensions regarding different legislation in different countries or regions or the influence of one country that detains the technology over others consuming it.  

\subsection{Fundamental concerns of GenAI}


GenAI models, particularly those based on deep learning architectures such as the GPT series, have shown remarkable capabilities in various SE tasks ranging from code generation to bug detection. However, their application in SE also brings a set of fundamental concerns. We summarized the concerns in the following RQs and elaborated on them further in this section:

\begin{enumerate}
    \item How all previous RQs can be addressed in an industry-level evaluation?
    \item How reliability and correctness of GenAI can be ensured for professional uses?
    \item How can GenAI output be explained and systematically integrated into current decision-making mechanisms?
    \item How sustainability concerns can be addressed for the application of GenAI?
\end{enumerate}
\textbf{Industry-level evaluation of GenAI.} 
 LLMs are seldom effective when used in a small task but can be highly effective as part of an overall software engineering process.
 The industry-level evaluation of GenAI technologies would be the next step to bridge the gap between small-sized empirical studies and real-world applications, ensuring that the promises made in research labs align with practical industry needs and challenges. Industry evaluations might reveal the tangible impacts, such as feasibility, usability, scalability, and cost-value tradeoff of adopting GenAI. Scalability is a critical factor for industry evaluation. Can the AI system perform well at scale, handling large volumes of data and complex tasks? Usability, user satisfaction, and user engagement are key aspects of evaluation. Moreover, research is desirable for understanding workflows, processes, and coordination paradigms so that GenAI can safely, efficiently and effectively reside. Moreover, in applications where AI interacts with users, the user experience is vital. 
 
\textbf{Reliability and correctness concern} The reliability and correctness of the generated code are pivotal. Unlike creative content generation, where occasional mistakes might be acceptable, in SE, even minor errors can have catastrophic consequences~\cite{se4ml}. However, there are challenges when rolling out improvements: improving a model's performance on certain tasks can unexpectedly affect its behavior in other tasks (e.g., fine-tuning additional data might enhance performance in one area but degrade it in another). The problem of hallucination has already been widely studied \cite{alkaissi_artificial_nodate,ma_scope_2023,azamfirei_large_2023,manakul_selfcheckgpt_2023,barrett_identifying_2023}. It will continue to be a topic of great interest, both within the software engineering community and in the wider computer science community. Moreover, the performance and behavior of both GPT-3.5 and GPT-4 vary greatly over time~\cite{gptchange}. For instance, GPT-4's March 2023 version had a better performance at identifying prime numbers compared to its June 2023 version, a relatively short period. This emphasizes the importance of continuous evaluation, especially given the lack of transparency regarding how models like ChatGPT are updated.

\textbf{Data availability} AI model relies heavily on a large amount of data for training and fine-tuning, posing challenges for private companies to adapt existing AI models to their work. The quality, diversity, and quantity of data directly affect the performance and generalizability of the models. Domain-specific data required for fine-tuning can be a bottleneck. Moreover, privacy and security are major concerns for companies and organizations adopting GenAI. Company data might be confidential, sensitive and personal. Concerning the adoption of GenAI in private companies can include concerns about data breaches, unauthorized access to confidential data, and the use of data without permission \cite{hamid_genaipabench_2023,kasneci_chatgpt_2023,manakul_selfcheckgpt_2023}. Regular audits of the data privacy and security measures will need to be in place to identify and address any potential vulnerabilities or areas for improvement.

\textbf{Explainability concern.} The ``black-box'' nature of these models poses interpretability challenges, making it difficult to reason about their decision-making process or to debug when things go awry~\cite{trustclassifier,trustpredictor}. Transparency in decision-making is crucial in SE, where understanding the rationale behind every line of code is often necessary.
Other ethical concerns also arise when discussing integrating GenAI tools in real-world environments. For instance, the dependence on training data means that these models can inadvertently introduce or perpetuate biases in the datasets they are trained on, leading to unfair or discriminatory software applications~\cite{datagenderbias}. These biases can be perpetuated if appropriate human oversight mechanisms, which require transparency, are not set in place. For instance, mechanisms to avoid automation bias are largely needed, especially for LLMs that usually make confident statements about their decisions. This could lead, in the future, as stated by the participants in the survey conducted by Cabrero-Daniel et al., to a potential loss of human expertise as tasks become automated and GenAI tools become common-use~\cite{perceivedtrust}, leading to a diminished ability to address or understand complex software challenges in the future manually.

\textbf{Sustainability concerns.}
As a general concern in sustainable software engineering \cite{amsel_toward_2011,penzenstadler_sustainability_2012,naumann_sustainable_2015}, the appearance of lots of LLMs daily raise a concern about their environmental impact, i.e. energy consumption \cite{chien_genai_2023,chien_reducing_2023,henderson_towards_2022}. Chiet et al. showed that ChatGPT-like services, inference dominates emissions, in one year producing 25 times the CO2 emissions of training GPT-3 \cite{chien_reducing_2023}. LLM capabilities have been associated with their size \cite{kaplan_scaling_2020}, resulting in the rapid growth of model size \cite{rae_scaling_2022}. While it has been suggested that the model performance depends not only on model size but also on the volume of training data \cite{hoffmann2022training}, the question of a triple-bottom-line model size that at the same time achieves performance, environmental and organizational objectives, remains unclear.


\section{Outlook and Conclusions}

The field of GenAI in SE is an important and fast-moving SE research area, currently presenting opportunities as well as open challenges. GenAI has the power to redefine the way software is designed, developed, and maintained. The innovative capabilities of AI-driven systems promise to streamline processes, boost efficiency, and open up new avenues for creative problem-solving. However, there are also common challenges in adopting GenAI and specific challenges for SE activities, i.e., requirements engineering, software implementation, quality assurance, and SE education. Our research agenda, outlined in this paper, stands as one of the pioneering initiatives to systematically reveal the state of research as of Oct 2023. Software Engineering is a large spectrum that includes technical activities but also user-centric, process-driven and managerial aspects. A comprehensive view should consider insights and methodologies from research about software project development, software project management, professional competencies, software engineering education, and software businesses. Therefore, collaboration with experts from these diverse disciplines is essential to harness the full potential of this research area.

In light of the evolving nature of GenAI research, significant work remains to establish GenAI in SE as a mature research area. Substantial theories must be developed to provide a solid theoretical underpinning for the practical applications. Technological advancements should be validated in a realistic project context. The adoption of GenAI in SE can also be combined with other active research areas, e.g., sustainability, trustworthy systems, and education.

We acknowledge that this research agenda might still miss some important SE aspects where GenAI can be applied. RQs are presented to illustrate significant concerns in each SE area and are not meant to be exhaustive. The research agenda is open to additions of new tracks, topics, and RQs by other researchers interested in the research area. Through this collective effort and commitment from a diverse community of researchers, we can strengthen the foundations of GenAI in SE and ensure its relevance in a rapidly changing technological landscape.
\clearpage
\textbf{Declaration of Generative AI and AI-assisted technologies in the writing process}
During the preparation of this work, we used ChatGPT version 3.5 and Grammarly services to edit and improve our writing, i.e. rephrasing sentences and fixing typos and language mistakes. After using this tool/service, we reviewed and edited the content as needed and take(s) full responsibility for the content of the publication.

\bibliographystyle{elsarticle-num}
\bibliography{main}
\end{document}